\documentstyle[a4,epsfig]{elsart}

\def\ltap{\;\raisebox{-.5ex}{\rlap{$\sim$}} \raisebox{.5ex}{$<$}\;}

\begin{document}
%}}} 

%{{{   titles, abstract etc
\rightline{SWAT/92}
\rightline{DESY 97-185}

\begin{frontmatter}

\title{Study of Lattice Correlation Functions at Small Times using the QCD
Sum Rules Continuum  Model}

\author{Chris Allton\thanksref{email1}}
\address{Department of Physics, University of Wales Swansea,
Singleton Park, \\ Swansea SA2 8PP, United Kingdom}

\author{Stefano Capitani\thanksref{email2}}
\address{DESY - Theory Group, Notkestrasse 85,\\
 D-22607 Hamburg, Germany}
\thanks[email1]{E-mail: c.allton@swansea.ac.uk}
\thanks[email2]{E-mail: stefano@mail.desy.de}

\begin{abstract}
In this paper we study the work of Leinweber by applying the Continuum
Model of QCD Sum Rules (QCDSR) to the analysis of (quenched) lattice
correlation functions. We expand upon his work in several areas: we
study meson states as well as baryons; we analyse data from several lattice
spacings; and we include data from the Sheikholeslami-Wohlert (clover)
improved action. We find that the QCDSR Continuum Model Ansatz can reproduce
the data, but only for non-physical values of its parameters. This
leads us to reject it as a model for hadronic correlation functions.

We study the non-relativistic quark model and conclude that it predicts
essentially the same form for the correlation function as the QCDSR
Continuum Model approach. Furthermore, because it doesn't have the
Continuum Model's restrictions on the parameters, the non-relativistic
quark model can be viewed as a successful Ansatz.

As well as studying the validity or otherwise of the QCDSR Continuum
Model approach, this paper defines 4-parameter fitting functions that
can be used to fit lattice data even for a time window close to the
source. These functions are shown to be an improvement over
2-exponential fits especially in the case of mesons. We encourage the
application of this approach to situations where the conventional
fitting procedures are problematic due to poor ground state dominance.

\end{abstract}

\end{frontmatter}
%}}} 

%{{{   introduction

\section{Introduction}

Lattice Gauge Theory and QCD Sum Rules (QCDSR) are two areas of
research which have been widely employed to deepen the understanding of
systems governed by the strong interactions. In this paper we combine
both methods following the pioneering work of Leinweber \cite{lw1,lw2}.
Utilising known results in QCD Sum Rules we apply them to Lattice Gauge
Theory in order to check that the two are compatible.

We begin by following the analysis of Refs.~\cite{lw1,lw2}. In this
work, the Continuum Model (often used in QCDSR analyses \cite{svz}) is
tested using lattice Monte Carlo data. The Continuum Model replaces
the discrete spectrum of excited states with a continuum of states
of a certain spectral density. This spectral density is calculated
using a Wilson OPE expansion of the relevant correlators in the small
and intermediate time separation regime. It has been argued that this
Continuum Model may be a good approximation to the actual situation of
a large set of discrete excited states \cite{svz,lw1}.

We extend the work of Refs. \cite{lw1,lw2} by including mesonic states.
We also apply our analysis to several $\beta$ values (corresponding to
different lattice spacings), and to two actions with different lattice
artefacts (namely the Wilson action \cite{wilson} and the ``clover''
Sheikholeslami-Wohlert (SW) improved action \cite{sw}). Thus we are
able to study the effects of lattice artefacts in the calculation. The
lattice data we use in this study is from the APE Collaboration.  As in
the analysis of \cite{lw1}, we use zero-momentum correlation functions.
At first sight the analysis suggests that the QCDSR Continuum Model is
in reasonably good agreement with the lattice data, confirming the
results of \cite{lw1}. However, this agreement is ambiguous since we
also find that the naive (non-relativistic) quark model predicts
essentially the same behaviour as the Continuum Model. Furthermore,
because we have data at different lattice spacings, we are able to test
the continuum scaling behaviour. We find that the parameter introduced
to take account of lattice distortions in the Continuum Model does not
scale towards its predicted value as the lattice spacing is reduced.
Thus we conclude from our analysis that there are reservations
regarding the validity of the QCDSR Continuum Model (but confirmation
of the quark model). We also test the relativistic quark model against
our data and find poorer agreement.

Through this study of the QCDSR Continuum Model, we have defined a
procedure which enables lattice correlation functions to be fitted at
small (Euclidean) time values. This can be usefully employed in the
case of correlation functions that have numerical problems in the
isolation of the ground state properties at large time separations. By
fitting the lattice correlation function data to a term representing
the ground state contribution, plus a term representing the
contributions of the excited states, we show that accurate ground state
parameters can be extracted even when the fitting window is very close
to the source. We explicitly show that these QCDSR-inspired fitting
functions reproduce the parameters of the ground state better than the
conventional, double-exponential fitting procedure.

Recently, some other researchers have also used Continuum Model ideas
to study lattice correlation functions \cite{chu,h}. They study
hadronic correlation function, but use configuration space (i.e.
point-to-point functions; see also \cite{shuryak}) rather than the
momentum space correlation functions analysed here.

This paper is organised as follows: in Sect.~2 details of the
lattice data used in our analysis are given; Sect.~3 describes the QCD
Sum Rules approach; in Sect.~4  results are presented for these fits;
in Sect.~5 we present results using the quark model; and finally, in
Sect.~6 there is a discussion and conclusion.
An earlier version of this work appeared in \cite{lat97}.

%}}} 

%{{{   lattice details
\section{Lattice Details}
\label{lat_details}

The lattice data used in this study is from the APE collaboration and
uses the quenched approximation. Specifically it is data which
has been used primarily in the study of weak matrix elements such as
$f_B$ and $B_K$ \cite{ape_sme0,ape_fbstat,ape_kbar}, and, more recently,
for light hadronic spectroscopy \cite{ape_light}. A list of the
simulation parameters is presented in Table \ref{tab_latt_params}. For
full details of the simulations, see the corresponding reference.
All errors in the results were obtained using the Jacknife
method with 10 configurations eliminated per cluster.

The lattices in this study span a range of $\beta$-values which
correspond to different lattice spacings, $a$.  In Table
\ref{tab_latt_params} and throughout this paper, $a$ is set from the
$K$ and $K^\ast$ masses following the approach described in 
\cite{ape_light}. This method avoids any chiral
extrapolation and uses data at the simulated values of the quark
masses, i.e. at around the strange mass. It is therefore preferable
than other methods, e.g. obtaining $a^{-1}$ from the $\rho$-mass etc.
Note that the lattice spacings are smaller than those used in
Refs.~\cite{lw1} and \cite{lw2} and so clearly lattice artefacts should
be smaller.

Typically each simulation was performed with three different values of
the hopping parameter, $\kappa$. Only one of these (for each
simulation) was chosen to be included in this study. This choice was
made so that the vector meson masses in physical units from each
simulation were equal to 1.1 $GeV$ within errors (see Table
\ref{tab_latt_params}). The one exception to this is the Wilson case at
$\beta=6.1$ where there was no $\kappa$ value available to match this
requirement.

%{{{   tab_latt_params
\begin{table}
\caption{\it{Simulation parameters used. The inverse lattice spacing
$a^{-1}$ was obtained from the $K$ and $K^\ast$ masses
\protect\cite{ape_light}.}}
\protect\label{tab_latt_params}

\begin{tabular}{lcccccccc}
\hline
Ref.     	&$\beta$& Action & Number of
                                       & Lattice        & $a^{-1}$ & $a$
                        &$\kappa$& $M_V$ \\
&&&Configs&Volume&[GeV]&[fm]&&[GeV]\\
\hline
%no60
\cite{ape_fbstat}&6.0   & Clover & 200 & $18^3\times64$	& 2.01(6)       & 0.098(3)
                        & 0.1425 & 1.10(3) \\
%no62
\cite{ape_sme0} & 6.2   & Clover & 200 & $18^3\times64$	& 3.0(3)	& 0.065(6)
                        & 0.14144& 1.14(11) \\
%6424_C
\cite{ape_fbstat}&6.4   & Clover & 400 & $24^3\times64$	& 3.9(2)	& 0.050(2)
                        & 0.1403 & 1.12(4) \\
\hline
%kbar
\cite{ape_kbar}	& 6.0	& Wilson & 200 & $18^3\times64$	& 2.27(4)	& 0.087(2)
                        & 0.1530 & 1.11(2) \\
%w61
\cite{ape_fbstat}&6.1   & Wilson & 230 & $18^3\times64$	& 2.7(2)	& 0.073(4)
                        & 0.1510 & 1.27(8) \\
%6424_W
\cite{ape_fbstat}&6.4   & Wilson & 400 & $24^3\times64$	& 4.1(2)	& 0.048(2)
                        & 0.1492 & 1.10(3) \\
\hline
\end{tabular}
\end{table}
%}}} 

The (zero-momentum) hadronic correlation function is defined for
mesons as
\begin{equation}
G_2(t) \equiv \sum_{\vec{x}} \langle 0|T\{J(\vec{x},t)
\overline{J}(\vec{0},0)\} |0 \rangle.
\label{eq:g2}
\end{equation}
While the definition of the meson correlators is standard, the
definition of the nucleon and delta correlation
functions is specific to the APE simulations, and is given by
\begin{equation}
 G_2(t) \equiv \sum_{\vec{x}} \mathrm{tr} \Bigg[
\frac{1+\gamma_4}{4}~\langle 0|T\{J(\vec{x},t)
\overline{J}(\vec{0},0)\} |0 \rangle \Bigg],
\label{eq:g2_ape}
\end{equation}
where the projector $(1+\gamma_4)$ onto positive parity states is
introduced. Another common definition for the nucleon
correlation function is the one used by the UKQCD
Collaboration (see for example \cite{h,ukqcd}). It is given by
\begin{equation}
 G_2(t) \equiv \sum_{\vec{x}} \mathrm{tr} \Bigg[
\frac{1}{4}~\langle 0|T\{J(\vec{x},t)
\overline{J}(\vec{0},0)\} \gamma_{\mu} x_{\mu} |0 \rangle \Bigg].
\label{eq:g2_ukqcd}
\end{equation}
Of specific interest to this work is the definitions of the
interpolating operators, $J$, defined below.
The channels analysed in this work are the nucleon, pseudoscalar
meson, (spatial component of the) vector meson and (temporal component
of the) axial meson. The other channels are included below for completeness.
\begin{description}
\item[Nucleon:] \ \ \
\begin{equation}
J(\vec{x},t) = \sum_\delta [ u^a(\vec{x},t) {\cal C} \gamma_5 d^b(\vec{x},t) ]
u_\delta^c(\vec{x},t) \epsilon_{abc}
\label{eq:J_nuc}
\end{equation}
\item[Delta:]\mbox{see \cite{ioffe}} \ \ \
\begin{equation}
J(\vec{x},t) = \sum_{\mu=2}^3 \sum_\delta [u^a(\vec{x},t) {\cal C} \gamma_{\mu}
u^b (\vec{x},t) ]  u_\delta^c(\vec{x},t) \epsilon_{abc}
\label{eq:J_del}
\end{equation}
\item[Scalar meson:] \ \ \
\begin{equation}
J(\vec{x},t) = \overline{u}^a(\vec{x},t) d^a(\vec{x},t)
\label{eq:J_sc}
\end{equation}
\item[Pseudoscalar meson:] \ \ \
\begin{equation}
J(\vec{x},t) = \overline{u}^a(\vec{x},t) \gamma_5 d^a(\vec{x},t)
\label{eq:J_ps}
\end{equation}
\item[Vector meson (spatial components):] \ \ \
\begin{equation}
J(\vec{x},t) = \frac{1}{3} \sum_i \overline{u}^a(\vec{x},t) \gamma_i d^a(\vec{x},t)
\label{eq:J_vi}
\end{equation}
\item[Vector meson (temporal component):] \ \ \
\begin{equation}
J(\vec{x},t) = \overline{u}^a(\vec{x},t) \gamma_4 d^a(\vec{x},t)
\label{eq:J_vt}
\end{equation}
\item[Axial meson (spatial components):] \ \ \
\begin{equation}
J(\vec{x},t) = \frac{1}{3} \sum_i \overline{u}^a(\vec{x},t) \gamma_5 \gamma_i d^a(\vec{x},t)
\label{eq:J_ai}
\end{equation}
\item[Axial meson (temporal component):] \ \ \
\begin{equation}
J(\vec{x},t) = \overline{u}^a(\vec{x},t) \gamma_5 \gamma_4
d^a(\vec{x},t) .
\label{eq:J_at}
\end{equation}
\end{description}

In these equations, the indices $a,b,c$ refer to colour, $\delta$ is a
spinorial index and ${\cal C} =  \gamma_4 \gamma_2$ is the charge
conjugation matrix.

All the baryonic correlation functions presented in this work have the
sum over $\vec{x}$ in Eq.~(\ref{eq:g2_ape}) replaced by a sum over
every 3rd lattice site in each of the three spatial directions. This
procedure, called ``thinning'', is due to the limitations imposed on
the calculation by the size of the APE memory. It introduces extra
states in the correlation functions which are not present when the
usual full sum over $\vec{x}$ it taken. These unwanted states have
large energy corresponding to large spatial momentums and do not affect
the conventional extraction of the ground state properties
\cite{ape_thin,ape_sme0,ape_light}. This is because at the (relatively
large) times where there is ground state dominance, the higher energy
states have decayed away. However, since the procedure presented in
this paper uses small time values, thinning does have an effect which
will be referred to in Sect.\ref{excited}.

%}}} 

%{{{   Continuum Model

\section{QCD (Sum Rules) Continuum Model}
\label{cont_model}

In this section we describe the QCDSR Continuum Model. The basic object
in this approach is the quark propagator in Euclidean space whose
first few polynomial terms in the
Wilson OPE expansion read \cite{lw1,chu} (in the coordinate gauge,
$x^{\mu}A_{\mu}=0$),
\begin{eqnarray} \nonumber
S^{aa'}_q & = &
   \frac{1}{2 \pi^2} \frac{\gamma \cdot x}{x^4}\delta^{aa'}
 + \frac{1}{4 \pi^2}\frac{m_q}{x^2}\delta^{aa'}
 - \frac{1}{8 \pi^2}\frac{m_q^2 \;\; \gamma \cdot x}{x^2}\delta^{aa'} \\
& & -\frac{1}{2^23} \langle : \overline{q}q : \rangle \delta^{aa'} +
\cdots.
\label{eq:qprop}
\end{eqnarray}
From Eqs.~(\ref{eq:g2},\ref{eq:g2_ape} \& \ref{eq:g2_ukqcd}),
the timesliced (three-momentum-projected)
two-point correlation function then has the OPE expansion
\begin{equation}
G^{OPE}_2(t)= \sum_{n=-\infty}^{n_0} \frac{1}{t^n} \cdot C_n
O_n  (m_q, \langle : \overline{q}q : \rangle )
\label{eq:expans},
\end{equation}
where $C_n$ is a numerical coefficient, $O_n$ is some function, and $n_0$ a
positive integer. For the nucleon and delta, this leading term turns
out to be $n_0=6$ for the APE definition Eqs.~(\ref{eq:g2_ape}).
However, different definitions of
the correlators lead to different QCDSR Continuum Model formulae, and if
we were to use the UKQCD definition,  Eq.~(\ref{eq:g2_ukqcd}), we would
have $n_0=5$ for both the nucleon and delta. For the mesons the
exponents of the leading terms are given by $n_0\le 3$.

It is now useful to express the timesliced correlation function,
$G_2(t)$, in the spectral representation
\begin{equation}
G_2(t)=\int_0^{\infty} \rho (s) \e^{-st} \d s.
\label{eq:spectr}
\end{equation}
The OPE expansion of the spectral density function, $\rho^{OPE}(s)$, is
calculated by means of a simple inverse Laplace transform of
Eq.~(\ref{eq:expans}),
\begin{equation}
\rho^{OPE}(s) = \sum_{n=1}^{n_0} \frac{s^{n-1}}{(n-1)!} \cdot C_n
O_n  (m_q, \langle : \overline{q}q : \rangle ),
\label{eq:rho}
\end{equation}
where $Re(t) > 0$. In the sum over $n$, we have kept only the positive
values of $n$ since we are interested in the leading terms as $t
\rightarrow 0$.

The QCD Continuum Model is introduced at this stage by setting a
threshold $s_0$ in the energy scale $s$ in
Eq.~(\ref{eq:spectr}), so that the excited states' contribution to
$G_2(t)$ is given only by the energies above $s_0$. Performing the
integral over $s$ we obtain,
\begin{eqnarray} \nonumber
G_2^{cont}(t) &\equiv& \int_{s_0}^{\infty} \rho^{OPE}(s) \e^{-st} \d s \\
&=& \e^{-s_0t} \sum_{n=1}^{n_0}
\sum_{k=0}^{n-1} \frac{1}{k!}\frac{s_0^k}{t^{n-k}}
\cdot C_n O_n  (m_q, \langle : \overline{q}q : \rangle ).
\label{eq:g2_cont}
\end{eqnarray}
Thus the continuum contribution to the correlation function has been
derived using QCDSR style considerations. The hope is that it correctly
models the discrete states in the spectrum above the ground state.

However, the full correlation function contains the ground state as well as the
above continuum contribution $G_2^{cont}$.
In the quenched approximation, the ground state contributes a delta
function to the full spectral density \cite{lw1,h}:
\begin{equation}
\rho(s) = \frac{Z}{2M} \delta(s-M) + \theta(s-s_0) \rho^{OPE}(s)
\end{equation}
where $\rho^{OPE}(s)$ is given by Eq.~(\ref{eq:rho}).
The full correlation function can then be written as (from
Eqs.~(\ref{eq:spectr} \& \ref{eq:g2_cont}))
\begin{eqnarray} \nonumber
G_2 (t)&=& \frac{Z}{2M} \e^{-Mt} + \xi
\int_{s_0}^{\infty} \rho^{OPE}(s) \e^{-st} \d s \\
       &=& \frac{Z}{2M} \e^{-Mt} + \xi \e^{-s_0t} \sum_{n=1}^{n_0}
\sum_{k=0}^{n-1} \frac{1}{k!}\frac{s_0^k}{t^{n-k}} \cdot C_n O_n
(m_q, \langle : \overline{q}q : \rangle )
\label{eq:model}
\end{eqnarray}

There are four parameters in this Ansatz for $G_2(t)$:
\begin{itemize}
\item $Z$ is the normalization of the ground state.
Note that in \cite{lw1,lw2} the parameter $\lambda_1$ was used.
$\lambda_1$ and $Z$ are related by $\lambda_1^2 = Z/(2M)$;
\item $M$ is the mass of the ground state particle;
\item $\xi$ is a new parameter introduced to normalise the contribution
of the excited states \cite{lw1}. In the continuum limit,
$\xi$ should be equal to one.
(We will discuss this in more detail in Sect.~\ref{excited});
\item $s_0$ is the continuum threshold which parametrises the onset
of the excited states.
\end{itemize}

We have developed some FORM codes able to calculate the coefficients
$C_n$ in the expansions in Eqs.~(\ref{eq:expans}) and (\ref{eq:model})
for  the operators in Eqs.~(\ref{eq:J_nuc} - \ref{eq:J_at}). The
relevant OPE expansions (i.e. the $G^{OPE}_2(t)$ in
Eq.~(\ref{eq:expans})) turn out to be:

\begin{description}
\item[Nucleon (APE data):]
\ \ \
\begin{displaymath}
G^{OPE}_2(t)=\frac{75}{256 \pi^4}\frac{1}{t^6}
+ \frac{21}{ 64 \pi^4}\frac{m_q  }{t^5}
- \frac{ 3}{ 16 \pi^4}\frac{m_q^2}{t^4}
\end{displaymath}
\begin{equation}
\hskip15mm - \frac{7}{32 \pi^2}\frac{\langle : \overline{q}q : \rangle}{t^3}
+ \cdots
\end{equation}
\item[Delta (APE data):]
\ \ \
\begin{displaymath}
G^{OPE}_2(t)=
  \frac{45}{32 \pi^4}\frac{1}{t^6}
+ \frac{15}{8 \pi^4}\frac{m_q}{t^5}
%+ \frac{15}{16 \pi^4}\frac{m_q^2}{t^4}
\end{displaymath}
\begin{equation}
\hskip15mm - \frac{15}{4 \pi^2}\frac{\langle : \overline{q}q : \rangle}{t^3}
+ \cdots
\end{equation}
\item[Scalar meson:]
\ \ \
\begin{equation}
G^{OPE}_2(t)=\frac{3}{4 \pi^2}\frac{1}{t^3}
- \frac{9}{4 \pi^2}\frac{m_q^2}{t}
+ \cdots
\end{equation}
\item[Pseudoscalar meson:]
\ \ \
%\newline
\begin{equation}
G^{OPE}_2(t)=-\frac{3}{4 \pi^2}\frac{1}{t^3}
+ \frac{3}{4 \pi^2}\frac{m_q^2}{t}
+ \cdots
\label{eq:ps_ope}
\end{equation}
\item[Vector meson (Spatial components):]
\ \ \
\begin{equation}
G^{OPE}_2(t) = - \frac{1}{2 \pi^2}\frac{1}{t^3}
+ \cdots
\label{eq:vi_ope}
\end{equation}
\item[Axial meson (Temporal components):]
\ \ \
\begin{equation}
G^{OPE}_2(t) =
  \frac{-3}{2 \pi^2}\frac{m_q^2}{t}
+ \cdots
\label{eq:axt_ope}
\end{equation}
\item[Axial meson (Spatial components):]
\ \ \
\begin{equation}
G^{OPE}_2(t) = \frac{1}{2 \pi^2}\frac{1}{t^3}
- \frac{3}{2 \pi^2}\frac{m_q^2}{t}
+ \cdots
\end{equation}
\end{description}

Note that for the vector meson, the leading term for the temporal
component is ${\cal O}(m_q^4)$. Since this corresponds to the first {\em
neglected} term in the OPE expansion for the quark propagator in
eq.(\ref{eq:qprop}), it is beyond the scope of this work and we
therefore do not consider this channel again.
Note also that there is no term ${\cal O}(m_q^2)$ for the spatial
component of the vector correlation function.

In the nucleon and delta case we have included for completeness the
contribution of the vacuum condensate. However, in practice, it is not
necessary to consider neither this term nor the $m_q^3/t^3$ term,
since they are numerically irrelevant. We note that Leinweber came to
the same  conclusion \cite{lw1}. It is important to note that for all
mesons listed there is no term appearing which is one order in $t$ smaller
than the leading term. i.e. the first nonzero correction, with respect
to the leading term, is of order $t^2$ and not $t$ as we would naively
expect. Similarly, the order $t^3$ term (with respect to the leading
term) is zero. The first appearance  of the vacuum condensate
$\langle : \overline{q}q : \rangle$ in the meson case is of order
$t^4$ relative to the leading term, rendering its contribution totally
insignificant.

We give here for the sake of completeness the correlation functions
corresponding to Eq.~(\ref{eq:model}), i.e. the $G_2(t)$ actually used
to fit the lattice data. They are:

\begin{description}
\item[Nucleon (APE data):]
\ \ \
\begin{eqnarray} \nonumber
G_2(t) &= \frac{Z}{2M} \e^{-Mt}& \\\nonumber
               &+ \xi \e^{-s_0t} \cdot &\Bigg[
  \frac{75}{256 \pi^4} \Bigg( \frac{1}{t^6}+\frac{s_0}{t^5}
 +\frac{1}{2}\frac{s_0^2}{t^4}+\frac{1}{6}\frac{s_0^3}{t^3}
 +\frac{1}{24}\frac{s_0^4}{t^2}+\frac{1}{120}\frac{s_0^5}{t}\Bigg) \\\nonumber
        &&+ \frac{21}{64 \pi^4} \;m_q \Bigg( \frac{1}{t^5}+\frac{s_0}{t^4}
 +\frac{1}{2}\frac{s_0^2}{t^3}+\frac{1}{6}\frac{s_0^3}{t^2}
 +\frac{1}{24}\frac{s_0^4}{t}\Bigg) \\
        &&- \frac{3}{16  \pi^4} \;m_q^2 \Bigg( \frac{1}{t^4}+\frac{s_0}{t^3}
 +\frac{1}{2}\frac{s_0^2}{t^2}+\frac{1}{6}\frac{s_0^3}{t}\Bigg)
 \Bigg]
\label{eq:nuc_fit}
\end{eqnarray}
\item[Delta (APE data):]
\ \ \
\begin{eqnarray} \nonumber
G_2(t) &= \frac{Z}{2M} \e^{-Mt}& \\\nonumber
        &+ \xi \e^{-s_0t} \cdot & \Bigg[
  \frac{45}{32 \pi^4} \Bigg( \frac{1}{t^6}+\frac{s_0}{t^5}
 +\frac{1}{2}\frac{s_0^2}{t^4}+\frac{1}{6}\frac{s_0^3}{t^3}
 +\frac{1}{24}\frac{s_0^4}{t^2}+\frac{1}{120}\frac{s_0^5}{t}\Bigg) \\\nonumber
        &&+ \frac{15}{8 \pi^4} \;m_q \Bigg( \frac{1}{t^5}+\frac{s_0}{t^4}
 +\frac{1}{2}\frac{s_0^2}{t^3}+\frac{1}{6}\frac{s_0^3}{t^2}
 +\frac{1}{24}\frac{s_0^4}{t}\Bigg)
%
%        &&+ \frac{15}{16 \pi^4} \;m_q^2 \Bigg( \frac{1}{t^4}+\frac{s_0}{t^3}
% +\frac{1}{2}\frac{s_0^2}{t^2}+\frac{1}{6}\frac{s_0^3}{t}\Bigg)
 \Bigg]
\label{eq:del_fit}
\end{eqnarray}
\item[Scalar mesons:]
\ \ \
\begin{equation}
G_2(t)= \frac{Z}{2M} \e^{-Mt} + \xi \e^{-s_0t} \cdot \Bigg[
  \frac{3}{4 \pi^2} \Bigg( \frac{1}{t^3}+\frac{s_0}{t^2}
 +\frac{1}{2}\frac{s_0^2}{t}\Bigg)
- \frac{9}{4 \pi^2}\frac{m_q^2}{t} \Bigg]
\label{eq:sca_fit}
\end{equation}
\item[Pseudoscalar mesons:]
\ \ \
\begin{equation}
G_2(t)= \frac{Z}{2M} \e^{-Mt} - \xi \e^{-s_0t} \cdot \Bigg[
  \frac{3}{4 \pi^2} \Bigg( \frac{1}{t^3}+\frac{s_0}{t^2}
 +\frac{1}{2}\frac{s_0^2}{t}\Bigg)
- \frac{3}{4 \pi^2}\frac{m_q^2}{t} \Bigg]
\label{eq:ps_fit}
\end{equation}
\item[Vector mesons (Spatial components):]
\ \ \
\begin{equation}
G_2(t)= \frac{Z}{2M} \e^{-Mt} - \xi \e^{-s_0t} \cdot \Bigg[
  \frac{1}{2 \pi^2} \Bigg( \frac{1}{t^3}+\frac{s_0}{t^2}
 +\frac{1}{2}\frac{s_0^2}{t}\Bigg) \Bigg]
\label{eq:vi_fit}
\end{equation}
\item[Axial mesons (Temporal components):]
\ \ \
\begin{equation}
G_2(t)= \frac{Z}{2M} \e^{-Mt} - \xi \e^{-s_0t} \cdot \Bigg[
 \frac{3}{2 \pi^2}\frac{m_q^2}{t} \Bigg] .
\label{eq:axt_fit}
\end{equation}
\item[Axial mesons (Spatial components):]
\ \ \
\begin{equation}
G_2(t)= \frac{Z}{2M} \e^{-Mt} + \xi \e^{-s_0t} \cdot \Bigg[
  \frac{1}{2 \pi^2} \Bigg( \frac{1}{t^3}+\frac{s_0}{t^2}
 +\frac{1}{2}\frac{s_0^2}{t}\Bigg)
- \frac{3}{2 \pi^2}\frac{m_q^2}{t} \Bigg] .
\label{eq:axi_fit}
\end{equation}
\end{description}

We have carried out several four-parameter fits to the nucleon,
pseudoscalar, vector (spatial components) and axial-vector (temporal
components) using Eqs.~(\ref{eq:nuc_fit}, \ref{eq:ps_fit},
\ref{eq:vi_fit} \& \ref{eq:axt_fit}) to check the ideas presented
in this paper. The results of these fits are described in detail in the
following Section.

One could also introduce a further parameter $\Lambda$ as upper limit
of integration over $s$ in Eq.~(\ref{eq:model}) to account for  the
lattice cut-off. Ref.~\cite{lw1} reports that this  does not make any
difference as long as $t$ is not taken less than two. For this reason
we restrict our fits to $t\ge2$.

%}}} 

%{{{   Fits to data
\section{Fits to data}
\label{qsr_fits}

%{{{   intro
Correlation functions from the simulations outlined in table
\ref{tab_latt_params} were fitted to the following three different
functional forms:

\begin{description}
\item[Conventional Single State Fit {\em (``1-exp'')}]
\ \ \
\begin{equation}
F(t) = \frac{Z}{2M} e^{-Mt}
\label{eq:single_exp}
\end{equation}
This is the usual functional form used in the study of lattice
correlation functions.
\item[QCD Continuum Model Fit {\em (``Cont'')}]
\ \ \
\vskip 1mm
\noindent
The fitting forms in Eqs.~(\ref{eq:nuc_fit}-\ref{eq:axi_fit}).
(i.e. those obtained from the O.P.E. plus the Continuum Model
assumption in Sect.~\ref{cont_model})
\item[Conventional Two State Fit {\em (``2-exp'')}]
\ \ \
\begin{equation}
F(t) = \frac{Z}{2M} e^{-Mt} + \frac{Z'}{2M'} e^{-M't}
\label{eq:2states}
\end{equation}
This functional form was chosen since it is traditionally used as the
generalisation of Eq.~(\ref{eq:single_exp}) when attempts are made to
include the effects of the higher mass state(s). It is an
alternative to the Continuum Model fits (i.e. it has the same number of
parameters) and the results of both fits will be directly compared.
\end{description}

Note that for the mesons, we have symmetrised the ground state
exponential in time for all three cases above by taking into
account the backward moving state in the fits.

The conventional single state fits are only valid for asymptotic
states. This limits their applicability to a time window where the
``effective mass'' is constant in time.\footnote{The effective mass is
defined as $\log (G_2(t)/G_2(t+1))$.} We use the results of these fits
(in the asymptotic region) as standard values of the ground state
fitting parameters $Z$ and $M$ with which to compare the results of the
other two fitting methods. A list of the parameters $Z$ and $M$ and the
$\chi^2$ obtained from these single state fits, is displayed in Tables
\ref{tab_proton},\ref{tab_p5},\ref{tab_vi} \& \ref{tab_at} in the rows
marked {\em ``1-exp''}. The four channels: nucleon and the pseudoscalar, vector
(spatial components) and axial (temporal component) mesons were analysed.

The time windows used in each of the fitting methods are
listed in Table \ref{tab_times} and are chosen so that the effective
masses are stable. The starting points in the windows are a function of
the $\beta$ value, as expected, since the lattice spacing varies with
$\beta$.
In the case of the QCD Continuum Model fits, we use time windows which
begin very close to the origin: $t=2-28$ for $\beta=6.0$; $t=3-28$ for
$\beta=6.1$ \& $6.2$; $t=4-28$ for $\beta=6.4$. (See Table
\ref{tab_times}.) The starting times for these windows are chosen with
the criterion that $t\ge2$ (see Sect.~\ref{cont_model}) and that they
are approximately equal in physical units for all
simulations. The results of these fits are displayed in Tables
\ref{tab_proton},\ref{tab_p5},\ref{tab_vi} \& \ref{tab_at} in the rows
marked {\em ``Cont''}.

So that a direct comparison can be performed, the same time windows
were used to perform conventional two state fits using
Eq.~(\ref{eq:2states}). The results of the fitting parameters from
these fits are also displayed in Tables \ref{tab_proton},\ref{tab_p5},
\ref{tab_vi} \& \ref{tab_at} in the rows labelled {\em ``2-exp''}.

We now discuss the results of these fitting methods in detail.

%}}} 

%{{{   Ground State Parameters

\subsection{Ground State Parameters}
\label{gnd_state}

Concentrating on the comparison of the ground state parameters, $Z$
and $M$, from the ``Cont'' fit with those from the ``1-exp'' fit, we
see good agreement for the mesonic states. This is particularly
encouraging because very small time values were included in the
fitting window.\footnote{Note also that it is well known that the
statistical errors in the fitted $Z$ and $M$ values typically
underestimate the real errors due to correlations in the data
\cite{ukawa}.} However, for the baryons, poorer agreement is
observed. In fact the agreement between the $Z$ and $M$ values from
the Continuum Model and ``1-exp'' fit becomes worse as $\beta$
increases (i.e. as the continuum limit is approached). Overall, these
fits suggest that the QCD Continuum Model is a reliable method of
extracting ground state information in the mesonic case, but that
there is some question about its validity for the baryons. Note that
Leinweber, who studied baryons only, found good agreement
\cite{lw1}. A possible explanation for this is that he used data at a
larger lattice spacing and with poorer statistics than those studied
here. Note also that the baryonic data is ``thinned'' which will
obviously affect the normalisation of the higher mass states. This
could conceivably also affect the ground state fitting parameters in the
``Cont'' fits.

The results for $Z$ and $M$ for the ``2-exp'' fits require careful
interpretation. Comparing the ``2-exp'' and ``1-exp'' fits we see that
the values for both $Z$ and $M$ for the ``2-exp'' fits are larger than
those in the ``1-exp'' fit {\em in every case}. This presumably means
that the introduction of a ``raw'' second exponential biases the
parameters of the asymptotic state. Also, it can be seen that in every
case except one (the Wilson $\beta=6.0$, Pseudoscalar meson) the
``Cont'' values for $Z$ and $M$ reproduce the ``standard values''
(i.e. the ``1-exp'' values) better than the ``2-exp'' do.

We can understand this bias in the case of the ``2-exp'' fit as
follows. The true set of states that contributes to the spectrum
includes the ground state and many excited states. In the case of the
``2-exp'' fit, all the excited states are modelled as a single state.
Clearly the values of $M'$ and $Z'$ from this fit will be some average
over all the excited states. This means that the $M'$ value will be
larger than the {\em first} excited state's mass. However, of the
excited states, it is this first excited state that is the most
significant correction to the ground state at intermediate values of
$t$ (i.e. at $t$ values large enough so that the excited states are no
longer dominant, and small enough so that the ground state also has not
yet dominated the correlation function). This means that the
correlation function for the ``2-exp'' fit will be too small in these
intermediate values of $t$. To adjust for this mismatch, the value of
the ground state's mass, $M$ (and $Z$ since they are correlated) will
be shifted slightly higher in the ``2-exp'' fit.

Obviously, in order to avoid this bias, correct account must be taken
of all the excited states in the spectrum. The Continuum Model is one
way of modelling these states, and thus should overcome at least part
of the above shortcomings of the ``2-exp'' fit. The agreement of the
ground state parameters, $Z$ and $M$, between the Continuum Model and
``1-exp'' fit confirms this hypothesis.

%}}} 

%{{{   Quality of the Fits

\subsection{Quality of the Fits}

\begin{figure}
\begin{center}
\mbox{\epsfig{file=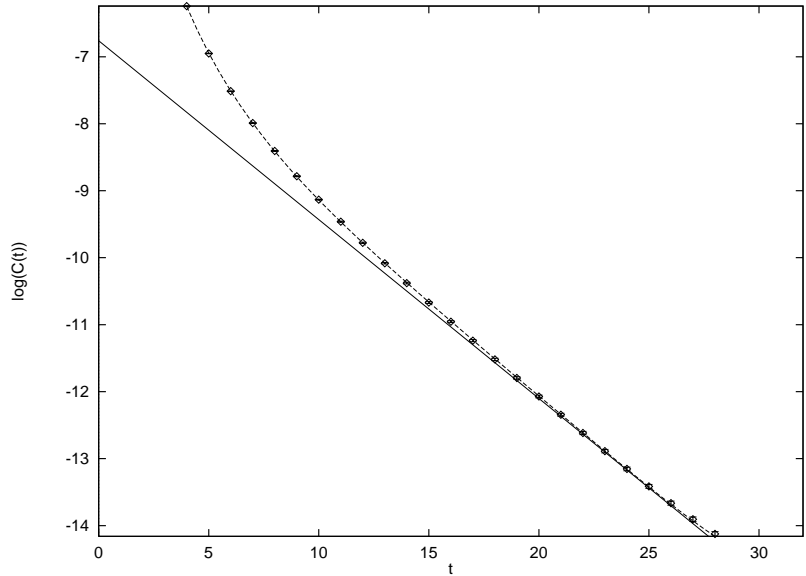, angle=270, width=11cm}}
\vskip 10mm
\caption{\it{Plot of the (natural) logarithm of the correlation function for the
$\beta=6.4$, Wilson, Vector meson data. The curve is the result of
the ``Cont'' fit and the straight line the result of the ``1-exp'' fit.
The ``2-exp'' fit is indistinguishable from the ``Cont'' fit on this plot.}}
\protect\label{fig_logC}
\end{center}
\end{figure}

Turning to the $\chi^2$ values, the first thing to note is that the
$\chi^2/\mbox{d.o.f.}$ for the ``Cont'' and ``2-exp'' fits can be as
much as 10 or 100. At first sight this is a worrying feature,
especially since the $\chi^2$ is calculated ignoring correlations
between timeslices. (The presence of correlations canonically lowers
the uncorrelated $\chi^2/\mbox{d.o.f.}$ to below unity.) However, the correlation
function data at small times has relatively tiny errors compared with
large times. (Typically the correlation function has statistical errors
of one part in a thousand at the beginning of the fitting window, and
around one part in ten at the end of the window.) Therefore any small
discrepancy between the fitting functions and the data at small times
leads to a very large contribution in the $\chi^2$. It is unreasonable
to expect that the small time behaviour of the correlation functions could
possibly be reproduced {\em at the level of these tiny statistics} by
any model containing only four parameters, no matter how well physically
motivated. It is also worth bearing in mind that the meson correlation
functions fall by from 3 to as much as 7 orders of magnitude in the
region fitted in the ``Cont'' and ``2-exp'' cases. In the case of the
baryons the correlation functions can fall by as much as 13 orders of
magnitude. Fig.~\ref{fig_logC} shows the (natural) logarithm of the
correlation function for the vector meson in the Wilson, $\beta=6.4$
case (which is a representative mesonic example). The plot shows the fits from
all three methods above. This figure illustrates the points made above;
the errors in the data for small times are clearly tiny.

We have argued that the absolute values of the $\chi^2/\mbox{d.o.f.}$
values might reasonably be expected to be large for the ``Cont'' and
``2-exp'' fits. However their {\em relative values} should correspond
to the relative quality of fits of the two methods.  From Tables
\ref{tab_proton},\ref{tab_p5},\ref{tab_vi} \& \ref{tab_at} we see
that the $\chi^2$ for the ``Cont'' fit are smaller than (or, at worse,
similar to) the ``2-exp'' fit in every case. In some cases
(particularly for the Pseudoscalar meson) the $\chi^2$ value for the
``Cont'' case is as much as an order of magnitude smaller than the
``2-exp'' case. This is again evidence that the ``Cont'' fitting
functions are an improvement over the traditional ``2-exp'' functions.

\begin{figure}
\begin{center}
\mbox{\epsfig{file=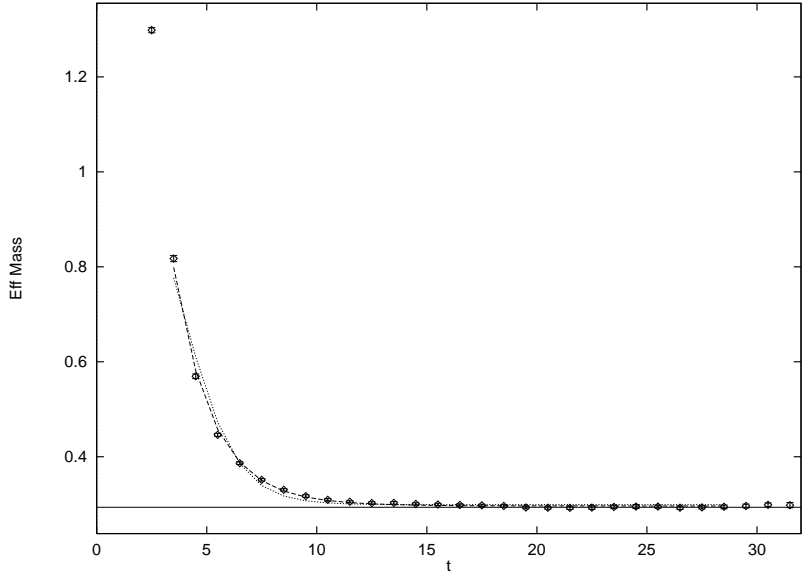, angle=270, width=11cm}}
\vskip 10mm
\caption{\it{Effective mass plot for the $\beta=6.2$, Clover, Pseudoscalar
data.
The values of $M$ from the fits to the ``1-exp'' is shown by a
horizontal solid line. The ``Cont'' is the dashed curve, and the
``2-exp'' is the dotted curve.
\protect\label{fig_effm1}}}
\end{center}
\end{figure}

\begin{figure}
\begin{center}
\mbox{\epsfig{file=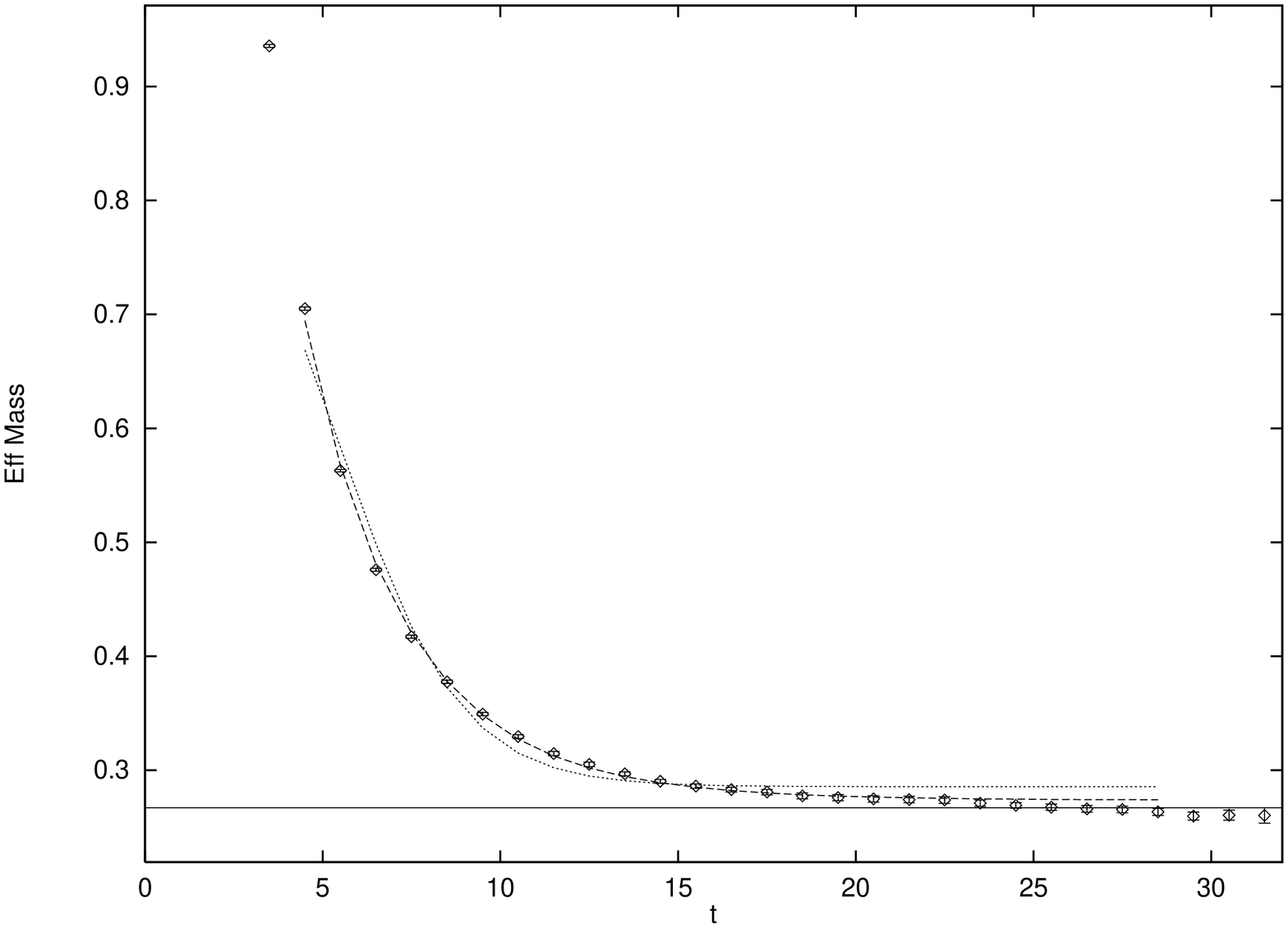, angle=270, width=11cm}}
\vskip 10mm
\caption{\it{Effective mass plot for the $\beta=6.4$, Wilson, Vector meson
data.
The values of $M$ from the fits to the ``1-exp'' is shown by a
horizontal solid line. The ``Cont'' is the dashed curve, and the
``2-exp'' is the dotted curve.}}
\protect\label{fig_effm2}
\end{center}
\end{figure}

\begin{figure}
\begin{center}
\mbox{\epsfig{file=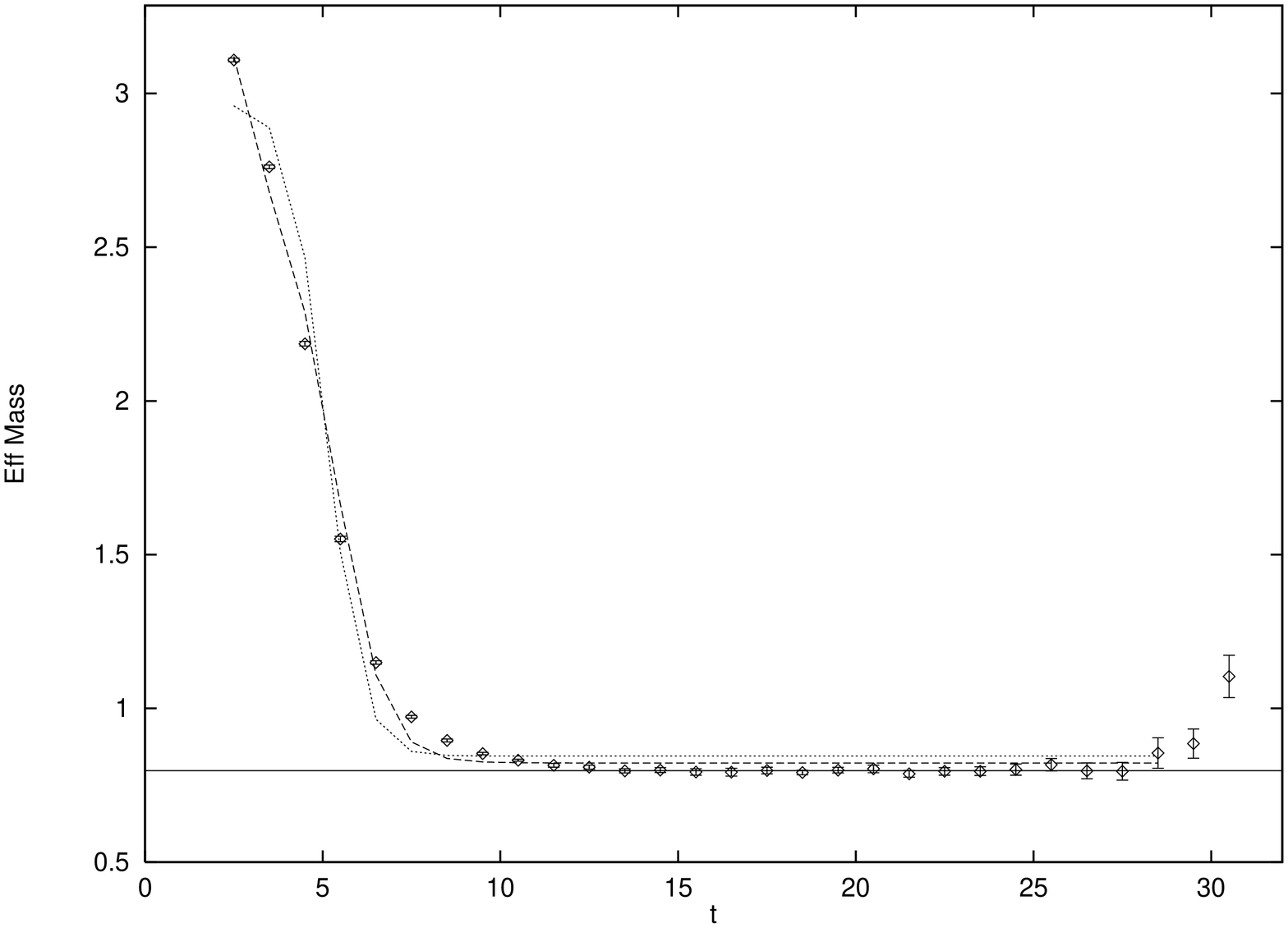, angle=270, width=11cm}}
\vskip 10mm
\caption{\it{Effective mass plot for the $\beta=6.0$, Wilson, Nucleon
data.
The values of $M$ from the fits to the ``1-exp'' is shown by a
horizontal solid line. The ``Cont'' is the dashed curve, and the
``2-exp'' is the dotted curve.}}
\protect\label{fig_effm3}
\end{center}
\end{figure}

As examples of the quality of the fits, Figs. \ref{fig_effm1},
\ref{fig_effm2} and \ref{fig_effm3} show the effective mass plots for
the Clover, $\beta=6.2$, Pseudoscalar case, the Wilson, $\beta=6.4$,
Vector meson case, and the Wilson, $\beta=6.0$, nucleon case
respectively. The central values of the effective mass from the three
methods: ``1-exp''; ``Cont''; and ``2-exp'' are shown. (Note that the
effective mass has been corrected for the effect of the backward moving
state and should therefore be constant at large times approaching the
middle of the lattice.) In these figures the ``2-exp'' fit can be seen
to give a reasonable reproduction of the Monte Carlo data, however, the
``Cont'' method performs a little better.  In particular the effect
discussed in Sect.~\ref{gnd_state} at intermediate time values can be
seen in the ``2-exp'' case. These three cases are a representative
sample of all the cases studied, and support the conclusions above
regarding the relative merits of the ``2-exp'' and ``Cont'' approaches.

\begin{figure}
\begin{center}
\mbox{\epsfig{file=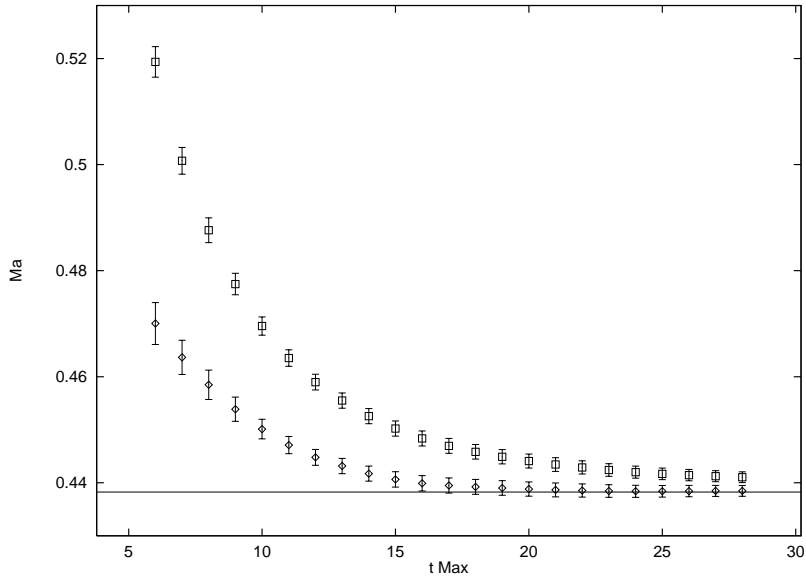, angle=270, width=11cm}}
\vskip 10mm
\caption{
\it{Results for the mass from the fit to the $\beta=6.0$, Clover,
pseudoscalar case for times $t=2 \rightarrow t_{MAX}$. 
The ``Cont'' fit results are shown as diamonds, and the ``2-exp'' case
as squares.
The ``1-exp'' fit for $t=12 \rightarrow 28$ is depicted by a horizontal solid line.}}
\protect\label{fig_effm4}
\end{center}
\end{figure}

\begin{figure}
\begin{center}
\mbox{\epsfig{file=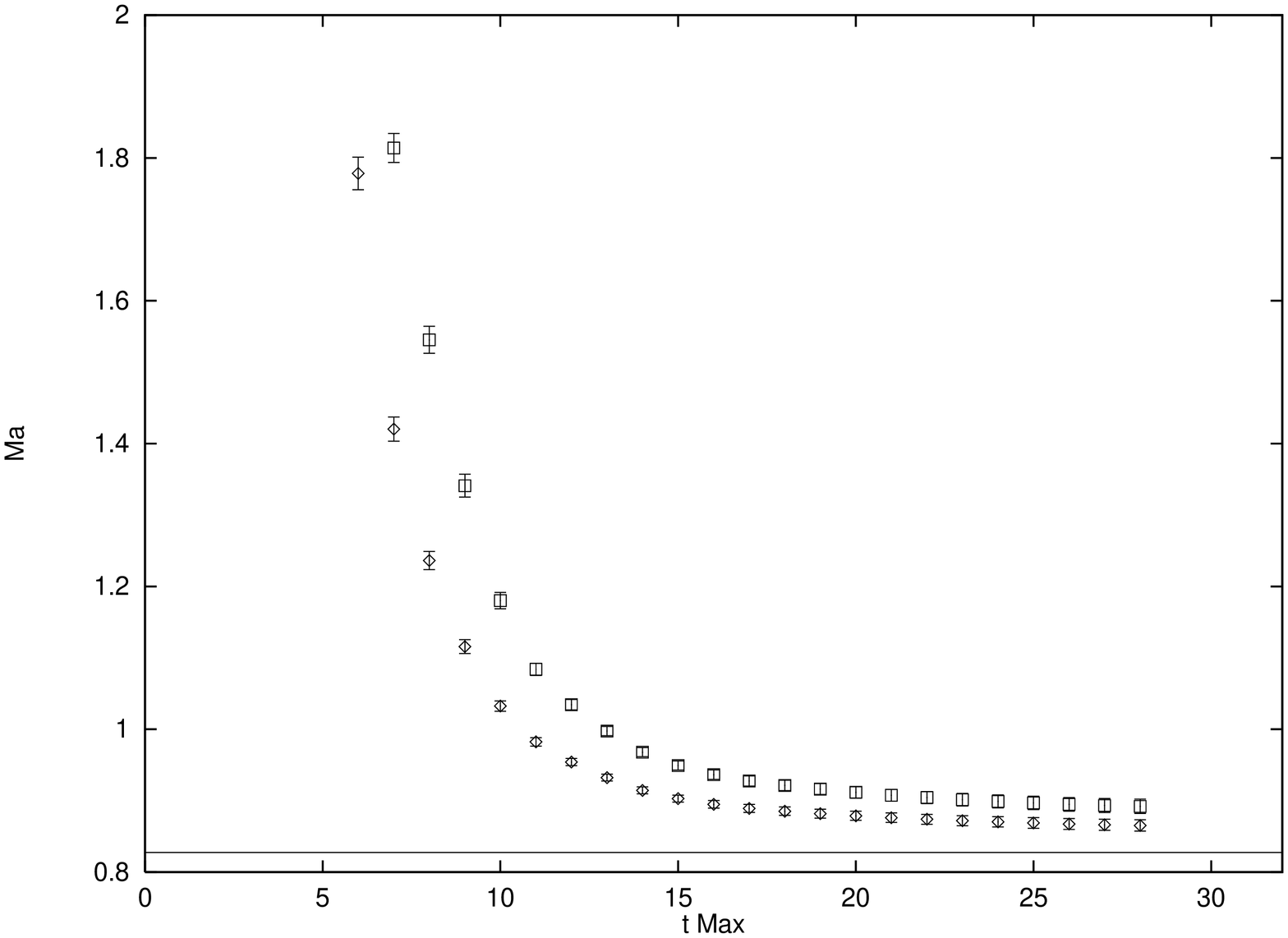, angle=270, width=11cm}}
\vskip 10mm
\caption{
\it{As in Figure \protect\ref{fig_effm4}, but for the Nucleon case.}}
\protect\label{fig_effm5}
\end{center}
\end{figure}

Further evidence that the ``Cont'' method is an improvement over the
``2-exp'' case can be seen in Figs. \ref{fig_effm4} and \ref{fig_effm5}.
The channels shown are for the pseudoscalar and nucleon respectively for
the Clover, $\beta=6.0$ simulation.
In these figures the mass, obtained from both the ``Cont'' and ``2-exp''
fits using $t=2 \rightarrow t_{MAX}$, is plotted against $t_{MAX}$.
(The ``1-exp'' fit using $t=12 \rightarrow 28$ is shown as the solid
horizontal line.)
As can be seen from the fits, the ``Cont'' method converges much faster
to the true mass value compared with the ``2-exp'' case.
Other channels show a similar behaviour.

%}}} 

%{{{   Parameters of the Excited State(s)

\subsection{Parameters of the Excited State(s)}
\label{excited}

Clearly it is not sufficient for the ``Cont'' fits to reproduce the
ground state parameters $Z$ and $M$, and to have a sensible $\chi^2$,
they must also give values for the continuum parameters, $s_0$ and
$\xi$, which are acceptable within the assumptions of the QCDSR
Continuum Model.

In Table~\ref{tab_s0} the values for $s_0$ (in $GeV$) are listed for
all the cases studied. These values must satisfy two criteria: (i)
since $s_0$ corresponds to a physical threshold, they should be
constant (in $GeV$) for each channel as $\beta$ is varied; and (ii)
they should be large enough to be in a region where perturbation theory
is valid. The second criterion is required so that the perturbative
expression for the quark propagator, Eq.~(\ref{eq:qprop}), is valid.
From Table~\ref{tab_s0}, it is clear that the $s_0$ values are roughly
constant in $\beta$ for the mesons, but not for the nucleon
(particularly the Clover case). Also, for
the pseudoscalar meson, the $s_0$ values are arguably too small ($\ltap
\;2 \;GeV$) for perturbation theory to be considered reliable. Thus
only the vector and axial mesons convincingly pass the criteria above.

We now turn to the values for $\xi$. Note that this parameter was
introduced to parametrise the distortions introduced in the lattice
formalism. It should therefore be unity in the continuum limit
\cite{lw1}. Table~\ref{tab_xi} displays the various values of $\xi$ from
the ``Cont'' fits.
Clearly, in the case of the axial meson, the $\xi$ values are not
``$\approx 1 + {\cal O}(a)$''.
The large size of these $\xi$ is because the leading term in the
Continuum part of $G_2(t)$ for this channel is ${\cal O}(m_q^2)$ (see
Eqs.~(\ref{eq:axt_fit}) and (\ref{eq:axt_ope})).
This implies that $\xi$ values several orders of magnitude larger than
the other channels (which have ${\cal O}(1)$ terms in $G_2(t)$) are
required.

In conclusion, the results of the continuum parameters, $s_0$ and
$\xi$, cast doubt over the applicability of the ``Cont'' model, since
the values obtained from the fits are not consistent with the
assumptions made in the QCDSR Continuum Model.

Note, it is not possible to compare directly the values for $\xi$ for
Wilson, nucleon data with those in Ref.~\cite{lw1}. This is because
the baryonic channels studied here have ``thinned'' correlation
functions (see the discussion in Sect.~\ref{lat_details}). This
affects the short-time region only, due to the presence of extra high
energy states, and thus conceivably alters the values of
$\xi$.
%}}} 

%{{{   Discussion on the Fits
\subsection{Discussion on the Fits}
\label{cont_discuss}

Summarising the subsections above, we see that the
QCDSR Continuum Model fitting functions reproduce the correct ground state
parameters, $Z$ and $M$, of the Monte Carlo data in the meson case.
The Continuum Model is not as successful however in the baryonic sector.
The {\em ``2-exp''} case does not appear to reproduce the ground state
parameters as effectively as the {\em ``Cont''} case for both baryons
and mesons. We presented an explanation for this finding in
Sect.~\ref{gnd_state}.
However, the values found for the continuum parameters $s_0$ and $\xi$
cast doubt over the applicability of the {\em ``Cont''} Ansatz, since
the values obtained are not consistent with the assumptions used in its
derivation. The conclusion therefore is to reject the QCDSR Continuum
Model as a way of modelling these lattice correlation functions.

However it is clear that, apart from taking on non-physical values of
its fitting parameters, the {\em ``Cont''} Ansatz does fairly
adequately reproduce the lattice correlation function data. Therefore
we require similar fitting Ans\"atze as those in the {\em ``Cont''}
case, but without the corresponding restrictions on the parameters
$s_0$ and $\xi$.

As we shall see in the next section, the naive non-relativistic quark
model predicts essentially the same functional form for the hadronic
correlators as in the QCDSR Continuum Model (for both the mesonic and
baryonic cases). Therefore it is a candidate model to use to fit the
data.

Before commencing the discussion on the quark model, we comment on the
results of the Wilson and clover actions. It is natural to ask if we
can observe any difference in the fitted parameters for these two
actions which might be a signal for improvement. Studying the $\chi^2$
values Tables \ref{tab_proton}, \ref{tab_p5}, \ref{tab_vi} \& \ref{tab_at}
it is difficult to see discernible differences between
the two actions.\footnote{It is possible that there is some difference
in the vector channel - see Table \ref{tab_vi}.} There are, however,
clear differences between the Continuum parameters in the two actions,
particularly for $\xi$ (see Table~\ref{tab_xi}). This is not surprising
since differences between the two actions are known to occur at small
times values (see \cite{h}). However, the interpretation of these
effects will require more work.
%}}} 

%{{{   table: proton
\begin{table}
\begin{center}

\begin{tabular}{llccc}
\hline
%\input tab_proton.tex
%{{{ table
 \multicolumn{5}{c}{\bf {                                            Nucleon}} \\
 \hline
 &&&&\\
 \multicolumn{5}{c}{\bf {                                     Clover Action}} \\
 &&&&\\
  $\beta$ & & 6.0 & 6.2 & 6.4 \\
 &&&&\\
 $Z$ 
  & 1-exp
  & $[     0.14 \pm    0.02 ] \times 10^{ -5 }$
  & $[     0.14 \pm    0.05 ] \times 10^{ -6 }$
  & $[     0.15 \pm    0.04 ] \times 10^{ -7 }$
  \\
  & Cont 
  & $[     0.24 \pm    0.02 ] \times 10^{ -5 }$
  & $[     0.30 \pm    0.03 ] \times 10^{ -6 }$
  & $[     0.48 \pm    0.02 ] \times 10^{ -7 }$
  \\
  & 2-exp
  & $[     0.34 \pm    0.04 ] \times 10^{ -5 }$
  & $[     0.33 \pm    0.03 ] \times 10^{ -6 }$
  & $[     0.50 \pm    0.02 ] \times 10^{ -7 }$
  \\
 $Ma$ 
  & 1-exp
  & $    0.828 \pm   0.009 $
  & $     0.59 \pm    0.02 $
  & $    0.438 \pm   0.009 $
  \\
  & Cont 
  & $    0.865 \pm   0.008 $
  & $    0.636 \pm   0.009 $
  & $    0.491 \pm   0.003 $
  \\
  & 2-exp
  & $     0.89 \pm    0.01 $
  & $    0.643 \pm   0.009 $
  & $    0.492 \pm   0.003 $
  \\
 $\chi^2$ 
  & 1-exp
  & $ (       1. \pm      1. ) /  15 $
  & $ (      0.7 \pm     0.8 ) /   9 $
  & $ (     0.02 \pm    0.05 ) /   3 $
  \\
  & Cont 
  & $ (    1000. \pm     70. ) /  23 $
  & $ (     310. \pm     60. ) /  22 $
  & $ (    1000. \pm     60. ) /  21 $
  \\
  & 2-exp
  & $ (    4000. \pm    100. ) /  23 $
  & $ (     550. \pm     80. ) /  22 $
  & $ (    1000. \pm     70. ) /  21 $
  \\
 \hline
 &&&&\\
 \multicolumn{5}{c}{\bf {                                     Wilson Action}} \\
 &&&&\\
  $\beta$ & & 6.0 & 6.1 & 6.4 \\
 &&&&\\
 $Z$ 
  & 1-exp
  & $[     0.16 \pm    0.02 ] \times 10^{ -5 }$
  & $[     0.61 \pm    0.07 ] \times 10^{ -6 }$
  & $[     0.23 \pm    0.04 ] \times 10^{ -7 }$
  \\
  & Cont 
  & $[     0.24 \pm    0.02 ] \times 10^{ -5 }$
  & $[    0.127 \pm   0.008 ] \times 10^{ -5 }$
  & $[     0.92 \pm    0.05 ] \times 10^{ -7 }$
  \\
  & 2-exp
  & $[     0.32 \pm    0.03 ] \times 10^{ -5 }$
  & $[    0.144 \pm   0.009 ] \times 10^{ -5 }$
  & $[    0.105 \pm   0.006 ] \times 10^{ -6 }$
  \\
 $Ma$ 
  & 1-exp
  & $    0.797 \pm   0.008 $
  & $    0.737 \pm   0.006 $
  & $    0.428 \pm   0.006 $
  \\
  & Cont 
  & $    0.823 \pm   0.007 $
  & $    0.776 \pm   0.005 $
  & $    0.495 \pm   0.004 $
  \\
  & 2-exp
  & $    0.845 \pm   0.008 $
  & $    0.784 \pm   0.005 $
  & $    0.503 \pm   0.005 $
  \\
 $\chi^2$ 
  & 1-exp
  & $ (      0.3 \pm     0.5 ) /  15 $
  & $ (      0.1 \pm     0.3 ) /  11 $
  & $ (     0.02 \pm    0.03 ) /   3 $
  \\
  & Cont 
  & $ (     220. \pm     40. ) /  23 $
  & $ (     660. \pm     60. ) /  22 $
  & $ (    2000. \pm     90. ) /  21 $
  \\
  & 2-exp
  & $ (    1000. \pm     80. ) /  23 $
  & $ (    1000. \pm     80. ) /  22 $
  & $ (    2000. \pm    100. ) /  21 $
  \\
%}}}
\hline
\end{tabular}

\caption{\it{Values for the fitting parameters for the nucleon.
The 1-exp, Cont and 2-exp refer to fits using
Eqs.~(\protect\ref{eq:single_exp}), (\protect\ref{eq:nuc_fit})
and (\protect\ref{eq:2states}) respectively.}}
\protect\label{tab_proton}

\end{center}
\end{table}
%}}} 
%{{{   table: p5
\begin{table}
\begin{center}

\begin{tabular}{llccc}
\hline
%\input tab_p5.tex
%{{{ table
 \multicolumn{5}{c}{\bf {                                Pseudoscalar Meson}} \\
 \hline
 &&&&\\
 \multicolumn{5}{c}{\bf {                                     Clover Action}} \\
 &&&&\\
  $\beta$ & & 6.0 & 6.2 & 6.4 \\
 &&&&\\
 $Z$ 
  & 1-exp
  & $   0.0328 \pm  0.0009 $
  & $[     0.80 \pm    0.05 ] \times 10^{ -2 }$
  & $[     0.27 \pm    0.01 ] \times 10^{ -2 }$
  \\
  & Cont 
  & $   0.0329 \pm  0.0007 $
  & $[     0.86 \pm    0.03 ] \times 10^{ -2 }$
  & $[    0.278 \pm   0.009 ] \times 10^{ -2 }$
  \\
  & 2-exp
  & $   0.0349 \pm  0.0007 $
  & $[     0.92 \pm    0.03 ] \times 10^{ -2 }$
  & $[    0.312 \pm   0.008 ] \times 10^{ -2 }$
  \\
 $Ma$ 
  & 1-exp
  & $    0.438 \pm   0.001 $
  & $    0.294 \pm   0.003 $
  & $    0.220 \pm   0.001 $
  \\
  & Cont 
  & $    0.438 \pm   0.001 $
  & $    0.296 \pm   0.002 $
  & $    0.220 \pm   0.001 $
  \\
  & 2-exp
  & $    0.441 \pm   0.001 $
  & $    0.299 \pm   0.002 $
  & $    0.224 \pm   0.001 $
  \\
 $\chi^2$ 
  & 1-exp
  & $ (      0.2 \pm     0.3 ) /  15 $
  & $ (     0.01 \pm    0.05 ) /   9 $
  & $ (    0.001 \pm   0.003 ) /   3 $
  \\
  & Cont 
  & $ (       3. \pm      1. ) /  23 $
  & $ (       3. \pm      2. ) /  22 $
  & $ (      1.1 \pm     0.5 ) /  21 $
  \\
  & 2-exp
  & $ (      48. \pm      5. ) /  23 $
  & $ (      18. \pm      4. ) /  22 $
  & $ (      21. \pm      4. ) /  21 $
  \\
 \hline
 &&&&\\
 \multicolumn{5}{c}{\bf {                                     Wilson Action}} \\
 &&&&\\
  $\beta$ & & 6.0 & 6.1 & 6.4 \\
 &&&&\\
 $Z$ 
  & 1-exp
  & $   0.0244 \pm  0.0007 $
  & $   0.0149 \pm  0.0005 $
  & $[    0.210 \pm   0.009 ] \times 10^{ -2 }$
  \\
  & Cont 
  & $   0.0226 \pm  0.0007 $
  & $   0.0144 \pm  0.0004 $
  & $[    0.200 \pm   0.009 ] \times 10^{ -2 }$
  \\
  & 2-exp
  & $   0.0253 \pm  0.0006 $
  & $   0.0155 \pm  0.0004 $
  & $[    0.236 \pm   0.005 ] \times 10^{ -2 }$
  \\
 $Ma$ 
  & 1-exp
  & $    0.423 \pm   0.001 $
  & $    0.404 \pm   0.001 $
  & $    0.205 \pm   0.001 $
  \\
  & Cont 
  & $    0.420 \pm   0.001 $
  & $    0.403 \pm   0.001 $
  & $    0.204 \pm   0.002 $
  \\
  & 2-exp
  & $    0.425 \pm   0.001 $
  & $    0.406 \pm   0.001 $
  & $    0.210 \pm   0.001 $
  \\
 $\chi^2$ 
  & 1-exp
  & $ (      0.3 \pm     0.4 ) /  15 $
  & $ (      0.1 \pm     0.2 ) /  11 $
  & $ (    0.002 \pm   0.004 ) /   3 $
  \\
  & Cont 
  & $ (      20. \pm      4. ) /  23 $
  & $ (       1. \pm      1. ) /  22 $
  & $ (      0.4 \pm     0.6 ) /  21 $
  \\
  & 2-exp
  & $ (      23. \pm      4. ) /  23 $
  & $ (      18. \pm      5. ) /  22 $
  & $ (      27. \pm      5. ) /  21 $
  \\
%}}}
\hline
\end{tabular}

\caption{\it{Values for the fitting parameters for the pseudoscalar
meson. The 1-exp, Cont and 2-exp refer to fits using
Eqs.~(\protect\ref{eq:single_exp}), (\protect\ref{eq:ps_fit})
and (\protect\ref{eq:2states}) respectively.}}
\protect\label{tab_p5}

\end{center}
\end{table}
%}}} 
%{{{   table: vi
\begin{table}
\begin{center}

\begin{tabular}{llccc}
\hline
%\input tab_vi.tex
%{{{ table
 \multicolumn{5}{c}{\bf {                 Vector Meson (Spatial Components)}} \\
 \hline
 &&&&\\
 \multicolumn{5}{c}{\bf {                                     Clover Action}} \\
 &&&&\\
  $\beta$ & & 6.0 & 6.2 & 6.4 \\
 &&&&\\
 $Z$ 
  & 1-exp
  & $[     0.93 \pm    0.06 ] \times 10^{ -2 }$
  & $[     0.18 \pm    0.03 ] \times 10^{ -2 }$
  & $[     0.51 \pm    0.04 ] \times 10^{ -3 }$
  \\
  & Cont 
  & $   0.0106 \pm  0.0003 $
  & $[     0.27 \pm    0.01 ] \times 10^{ -2 }$
  & $[     0.77 \pm    0.03 ] \times 10^{ -3 }$
  \\
  & 2-exp
  & $   0.0121 \pm  0.0004 $
  & $[    0.300 \pm   0.010 ] \times 10^{ -2 }$
  & $[     0.94 \pm    0.03 ] \times 10^{ -3 }$
  \\
 $Ma$ 
  & 1-exp
  & $    0.548 \pm   0.004 $
  & $    0.378 \pm   0.006 $
  & $    0.285 \pm   0.003 $
  \\
  & Cont 
  & $    0.556 \pm   0.003 $
  & $    0.397 \pm   0.003 $
  & $    0.300 \pm   0.002 $
  \\
  & 2-exp
  & $    0.565 \pm   0.003 $
  & $    0.403 \pm   0.003 $
  & $    0.309 \pm   0.002 $
  \\
 $\chi^2$ 
  & 1-exp
  & $ (       1. \pm      1. ) /  15 $
  & $ (      0.4 \pm     0.6 ) /   9 $
  & $ (     0.02 \pm    0.02 ) /   3 $
  \\
  & Cont 
  & $ (     130. \pm     30. ) /  23 $
  & $ (     120. \pm     20. ) /  22 $
  & $ (      91. \pm      2. ) /  21 $
  \\
  & 2-exp
  & $ (     610. \pm     60. ) /  23 $
  & $ (     270. \pm     40. ) /  22 $
  & $ (     290. \pm     30. ) /  21 $
  \\
 \hline
 &&&&\\
 \multicolumn{5}{c}{\bf {                                     Wilson Action}} \\
 &&&&\\
  $\beta$ & & 6.0 & 6.1 & 6.4 \\
 &&&&\\
 $Z$ 
  & 1-exp
  & $   0.0102 \pm  0.0004 $
  & $[     0.57 \pm    0.03 ] \times 10^{ -2 }$
  & $[     0.62 \pm    0.04 ] \times 10^{ -3 }$
  \\
  & Cont 
  & $[     0.95 \pm    0.03 ] \times 10^{ -2 }$
  & $[     0.59 \pm    0.02 ] \times 10^{ -2 }$
  & $[     0.74 \pm    0.03 ] \times 10^{ -3 }$
  \\
  & 2-exp
  & $   0.0120 \pm  0.0003 $
  & $[     0.68 \pm    0.02 ] \times 10^{ -2 }$
  & $[     0.99 \pm    0.03 ] \times 10^{ -3 }$
  \\
 $Ma$ 
  & 1-exp
  & $    0.508 \pm   0.003 $
  & $    0.466 \pm   0.002 $
  & $    0.267 \pm   0.003 $
  \\
  & Cont 
  & $    0.505 \pm   0.002 $
  & $    0.468 \pm   0.002 $
  & $    0.274 \pm   0.002 $
  \\
  & 2-exp
  & $    0.517 \pm   0.002 $
  & $    0.474 \pm   0.002 $
  & $    0.286 \pm   0.002 $
  \\
 $\chi^2$ 
  & 1-exp
  & $ (      0.3 \pm     0.6 ) /  15 $
  & $ (      0.4 \pm     0.6 ) /  11 $
  & $ (    0.008 \pm   0.009 ) /   3 $
  \\
  & Cont 
  & $ (      20. \pm      5. ) /  23 $
  & $ (       8. \pm      5. ) /  22 $
  & $ (      11. \pm      4. ) /  21 $
  \\
  & 2-exp
  & $ (     200. \pm     20. ) /  23 $
  & $ (     110. \pm     20. ) /  22 $
  & $ (     180. \pm     20. ) /  21 $
  \\
%}}}
\hline
\end{tabular}

\caption{\it{Values for the fitting parameters for the (spatial components
of the) vector
meson. The 1-exp, Cont and 2-exp refer to fits using
Eqs.~(\protect\ref{eq:single_exp}), (\protect\ref{eq:vi_fit})
and (\protect\ref{eq:2states}) respectively.}}
\protect\label{tab_vi}

\end{center}
\end{table}
%}}} 
%{{{   table: a0
\begin{table}
\begin{center}

\begin{tabular}{llccc}
\hline
%\input tab_a0.tex
%{{{ table
 \multicolumn{5}{c}{\bf {          Axial-Vector Meson (Temporal Components)}} \\
 \hline
 &&&&\\
 \multicolumn{5}{c}{\bf {                                     Clover Action}} \\
 &&&&\\
  $\beta$ & & 6.0 & 6.2 & 6.4 \\
 &&&&\\
 $Z$ 
  & 1-exp
  & $[    0.171 \pm   0.008 ] \times 10^{ -2 }$
  & $[     0.37 \pm    0.03 ] \times 10^{ -3 }$
  & $[    0.111 \pm   0.009 ] \times 10^{ -3 }$
  \\
  & Cont 
  & $[    0.179 \pm   0.006 ] \times 10^{ -2 }$
  & $[     0.38 \pm    0.02 ] \times 10^{ -3 }$
  & $[    0.118 \pm   0.004 ] \times 10^{ -3 }$
  \\
  & 2-exp
  & $[    0.180 \pm   0.006 ] \times 10^{ -2 }$
  & $[     0.38 \pm    0.02 ] \times 10^{ -3 }$
  & $[    0.118 \pm   0.004 ] \times 10^{ -3 }$
  \\
 $Ma$ 
  & 1-exp
  & $    0.437 \pm   0.002 $
  & $    0.293 \pm   0.004 $
  & $    0.217 \pm   0.003 $
  \\
  & Cont 
  & $    0.440 \pm   0.002 $
  & $    0.295 \pm   0.004 $
  & $    0.219 \pm   0.002 $
  \\
  & 2-exp
  & $    0.440 \pm   0.002 $
  & $    0.295 \pm   0.004 $
  & $    0.219 \pm   0.002 $
  \\
 $\chi^2$ 
  & 1-exp
  & $ (      0.3 \pm     0.2 ) /  15 $
  & $ (     0.06 \pm    0.06 ) /   9 $
  & $ (     0.02 \pm    0.02 ) /   3 $
  \\
  & Cont 
  & $ (      10. \pm      3. ) /  23 $
  & $ (      1.2 \pm     0.8 ) /  22 $
  & $ (       2. \pm      1. ) /  21 $
  \\
  & 2-exp
  & $ (      15. \pm      4. ) /  23 $
  & $ (      1.5 \pm     0.9 ) /  22 $
  & $ (       2. \pm      1. ) /  21 $
  \\
 \hline
 &&&&\\
 \multicolumn{5}{c}{\bf {                                     Wilson Action}} \\
 &&&&\\
  $\beta$ & & 6.0 & 6.1 & 6.4 \\
 &&&&\\
 $Z$ 
  & 1-exp
  & $[     0.23 \pm    0.01 ] \times 10^{ -2 }$
  & $[    0.159 \pm   0.006 ] \times 10^{ -2 }$
  & $[    0.126 \pm   0.009 ] \times 10^{ -3 }$
  \\
  & Cont 
  & $[    0.234 \pm   0.007 ] \times 10^{ -2 }$
  & $[    0.170 \pm   0.004 ] \times 10^{ -2 }$
  & $[    0.140 \pm   0.005 ] \times 10^{ -3 }$
  \\
  & 2-exp
  & $[    0.236 \pm   0.006 ] \times 10^{ -2 }$
  & $[    0.171 \pm   0.004 ] \times 10^{ -2 }$
  & $[    0.140 \pm   0.005 ] \times 10^{ -3 }$
  \\
 $Ma$ 
  & 1-exp
  & $    0.422 \pm   0.002 $
  & $    0.403 \pm   0.002 $
  & $    0.201 \pm   0.002 $
  \\
  & Cont 
  & $    0.423 \pm   0.002 $
  & $    0.406 \pm   0.002 $
  & $    0.205 \pm   0.001 $
  \\
  & 2-exp
  & $    0.423 \pm   0.002 $
  & $    0.406 \pm   0.002 $
  & $    0.205 \pm   0.001 $
  \\
 $\chi^2$ 
  & 1-exp
  & $ (      0.5 \pm     0.5 ) /  15 $
  & $ (      0.8 \pm     0.5 ) /  11 $
  & $ (     0.01 \pm    0.01 ) /   3 $
  \\
  & Cont 
  & $ (       2. \pm      2. ) /  23 $
  & $ (       5. \pm      3. ) /  22 $
  & $ (      0.8 \pm     0.9 ) /  21 $
  \\
  & 2-exp
  & $ (       4. \pm      2. ) /  23 $
  & $ (       6. \pm      3. ) /  22 $
  & $ (      0.9 \pm     1.0 ) /  21 $
  \\
%}}}
\hline
\end{tabular}

\caption{\it{Values for the fitting parameters for the (temporal components
of the) axial
meson. The 1-exp, Cont and 2-exp refer to fits using
Eqs.~(\protect\ref{eq:single_exp}), (\protect\ref{eq:axt_fit})
and (\protect\ref{eq:2states}) respectively.}}
\protect\label{tab_at}

\end{center}
\end{table}
%}}} 

%{{{   table: times
\begin{table}
\begin{center}

\begin{tabular}{lccc}
 \hline
 &&&\\
 \multicolumn{4}{c}{\bf {                                     Clover Action}} \\
 &&&\\
  $\beta$ & 6.0   & 6.2   & 6.4   \\
 &&&\\
  1-exp   & 12-28 & 18-28 & 24-28 \\
  Cont    & 2-28  & 3-28  & 4-28  \\
  2-exp   & 2-28  & 3-28  & 4-28  \\
 &&&\\
 \hline
 &&&\\
 \multicolumn{4}{c}{\bf {                                     Wilson Action}} \\
 &&&\\
  $\beta$ & 6.0   & 6.1   & 6.4   \\
 &&&\\
  1-exp   & 12-28 & 16-28 & 24-28 \\
  Cont    & 2-28  & 3-28  & 4-28  \\
  2-exp   & 2-28  & 3-28  & 4-28  \\
 &&&\\
 \hline
\end{tabular}

\caption{\it{The time windows used in each of the fitting methods.}}
\protect\label{tab_times}

\end{center}
\end{table}
%}}} 

%{{{   table: S_0

\begin{table}
\begin{center}

\begin{tabular}{llccc}
\hline
&&&&\\
\multicolumn{5}{c}{\bf {$s_0 \;\;\; [GeV]$}} \\
%{{{ table
 &&&&\\
 \hline
 &&&&\\
 \multicolumn{5}{c}{\bf {                                     Clover Action}} \\
 &&&&\\
  $\beta$ & & 6.0 & 6.2 & 6.4 \\
 &&&&\\
 \bf {                                            Nucleon} & 
  & $      5.8 \pm     0.2 $
  & $      8.5 \pm     0.8 $
  & $     10.1 \pm     0.5 $
  \\
 \bf {                                Pseudoscalar Meson} & 
  & $     1.87 \pm    0.06 $
  & $      2.0 \pm     0.2 $
  & $      1.6 \pm     0.1 $
  \\
 \bf {                 Vector Meson} & 
  & $     2.57 \pm    0.09 $
  & $      2.9 \pm     0.2 $
  & $      2.4 \pm     0.1 $
  \\
 \bf {(Spatial Components)} & &&\\
 \bf {          Axial-Vector Meson} & 
  & $      4.1 \pm     0.2 $
  & $      5.0 \pm     0.4 $
  & $      5.3 \pm     0.5 $
  \\
 \bf {(Temporal Components)} & &&\\
 &&&&\\
 \hline
 &&&&\\
 \multicolumn{5}{c}{\bf {                                     Wilson Action}} \\
 &&&&\\
  $\beta$ & & 6.0 & 6.1 & 6.4 \\
 &&&&\\
 \bf {                                            Nucleon} & 
  & $      5.0 \pm     0.1 $
  & $      5.9 \pm     0.4 $
  & $      6.9 \pm     0.3 $
  \\
 \bf {                                Pseudoscalar Meson} & 
  & $     1.34 \pm    0.05 $
  & $      1.7 \pm     0.1 $
  & $      1.4 \pm     0.1 $
  \\
 \bf {                 Vector Meson} & 
  & $     1.77 \pm    0.04 $
  & $      2.0 \pm     0.1 $
  & $     1.97 \pm    0.08 $
  \\
 \bf {(Spatial Components)} & &&\\
 \bf {          Axial-Vector Meson} & 
  & $      3.5 \pm     0.1 $
  & $      3.5 \pm     0.3 $
  & $      4.2 \pm     0.5 $
  \\
 \bf {(Temporal Components)} & &&\\
 &&&&\\
 \hline
%}}}
\end{tabular}

\caption{\it{Values for the fitting parameter $s_0$ in $GeV$ from the
``Cont'' fit.}}
\protect\label{tab_s0}

\end{center}
\end{table}

%}}} 
%{{{   table: xi
\begin{table}
\begin{center}

\begin{tabular}{llccc}
\hline
 &&&&\\
\multicolumn{5}{c}{\bf {$\xi$}} \\
%{{{ table
 &&&&\\
 \hline
 &&&&\\
 \multicolumn{5}{c}{\bf {                                     Clover Action}} \\
 &&&&\\
  $\beta$ & & 6.0 & 6.2 & 6.4 \\
 &&&&\\
 \bf {                                            Nucleon} & 
  & $     72.4 \pm     0.5 $
  & $     49.7 \pm     0.6 $
  & $     22.6 \pm     0.4 $
  \\
 \bf {                                Pseudoscalar Meson} & 
  & $     9.49 \pm    0.07 $
  & $      6.8 \pm     0.1 $
  & $     4.87 \pm    0.07 $
  \\
 \bf {                 Vector Meson} & 
  & $     9.41 \pm    0.07 $
  & $     5.70 \pm    0.10 $
  & $     2.85 \pm    0.04 $
  \\
 \bf {(Spatial Components)} & &&\\
 \bf {          Axial-Vector Meson} & 
  & $[     1.54 \pm    0.06 ] \times 10^{  3 }$
  & $[      1.3 \pm     0.2 ] \times 10^{  3 }$
  & $[       6. \pm      2. ] \times 10^{  2 }$
  \\
 \bf {(Temporal Components)} & &&\\
 &&&&\\
 \hline
 &&&&\\
 \multicolumn{5}{c}{\bf {                                     Wilson Action}} \\
 &&&&\\
  $\beta$ & & 6.0 & 6.1 & 6.4 \\
 &&&&\\
 \bf {                                            Nucleon} & 
  & $     6.73 \pm    0.03 $
  & $     5.58 \pm    0.06 $
  & $     1.82 \pm    0.04 $
  \\
 \bf {                                Pseudoscalar Meson} & 
  & $     3.82 \pm    0.01 $
  & $     4.03 \pm    0.06 $
  & $     3.74 \pm    0.04 $
  \\
 \bf {                 Vector Meson} & 
  & $     3.40 \pm    0.02 $
  & $     3.28 \pm    0.04 $
  & $     2.66 \pm    0.02 $
  \\
 \bf {(Spatial Components)} & &&\\
 \bf {          Axial-Vector Meson} & 
  & $     126. \pm      8. $
  & $      56. \pm      2. $
  & $      94. \pm      8. $
  \\
 \bf {(Temporal Components)} & &&\\
 &&&&\\
 \hline
%}}}
\end{tabular}

\caption{\it{Values for the fitting parameter $\xi$ from the
``Cont'' fit.}}
\protect\label{tab_xi}

\end{center}
\end{table}
%}}} 

%{{{   fig_logC
%}}} 

%{{{   fig_effm1
%}}} 
%{{{   fig_effm2
%}}} 
%{{{   fig_effm3
%}}} 

%{{{   fig_effm4

%}}} 
%{{{   fig_effm5
%}}} 

%}}} 

%{{{   quark models

\section{Quark Models}
\label{qm_fits}

%{{{   non-rel quark model

\subsection{Non-relativistic Quark Model}

The quark model has had considerable success in reproducing many of the
features of the hadronic spectrum \cite{qm}. It is thus natural to
check its predictions for the hadronic correlators with the lattice
Monte Carlo data. We present firstly the derivation of the hadronic
correlators for the non-relativistic quark model, and then the
relativistic case.

The aim of this exercise is to determine the density of states function
$\rho(s)$ which defines the number of states per unit energy range.
Taking the quarks as on-shell, there are 3 degrees of freedom available
for each quark in the hadron: one per spatial momentum component.
Therefore in the case of mesons there are naively 6 degrees of freedom,
with 9 degrees of freedom for baryons. However, there are 4 constraint
equations. Three correspond to demanding that the total 3-momentum is
fixed to the hadron's momentum (which is zero in the correlators we are
studying), and one which corresponds to demanding that the total kinetic
energy is fixed to $s$. This leaves 2 degrees of freedom for the mesons
and 5 degrees of freedom for the baryons. This simple, non-relativistic
analysis leads to the following form for the spectral density.
\begin{eqnarray}
&\mbox{For mesons: \hskip 5mm} & \rho_{nrqm}(s) \sim s^2 \\
&\mbox{For baryons: \hskip 5mm}& \rho_{nrqm}(s) \sim s^5
\end{eqnarray}
The above continuum-like behaviour for $\rho(s)$ is clearly most
appropriate for large $s$ where the quarks are approximately free.
Therefore the actual density of states can be approximated by
\begin{equation}
\rho(s) = \frac{Z}{2M} \; \delta(s-M) + \theta(s-s_0) \rho_{nrqm}(s),
\end{equation}
where the delta-function represents the ground state, and the non-relativistic
quark model result is used for the continuum (beginning at the threshold
energy $s_0$). Using this definition of
$\rho(s)$ in the spectral equation (Eq.~(\ref{eq:spectr})) we obtain
\vskip 1mm
for mesons:
\begin{equation}
G_2(t) = \frac{Z}{2M} e^{-Mt} + K \left(
\frac{1}{t^{3}} + \frac{s_0}{t^2} + \frac{s_0^2}{2t}
\right) e^{-s_0 t},
\label{eq:nrqm_mes}
\end{equation}
\vskip 1mm
and for baryons:
\begin{equation}
G_2(t) = \frac{Z}{2M} e^{-Mt} + K' \left(
\frac{1}{t^{6}} + \frac{s_0}{t^5} + \frac{s_0^2}{2\;t^4} +
\frac{s_0^3}{3!\; t^3} + \frac{s_0^4}{4!\; t^2} + \frac{s_0^5}{5!\; t}
\right) e^{-s_0 t},
\label{eq:nrqm_bar}
\end{equation}
where $K$ and $K'$ are some numerical constants. Note that
Eqs.~(\ref{eq:nrqm_mes}\&\ref{eq:nrqm_bar}) are identical to
Eqs.~(\ref{eq:nuc_fit}-\ref{eq:axi_fit})
in Sect.~\ref{cont_model} with $m_q=0$.
Since numerically the terms ${\cal O}(m_q)$ and ${\cal O}(m_q^2)$ are
insignificant, we have shown that the non-relativistic quark model predicts
essentially the same functional form as the QCD Sum Rules Continuum
Model for these channels.

Thus we have achieved the aim raised in Sect.~\ref{cont_discuss}: we
have found an Ansatz which reproduces the data better than the
``2-exp'' fits, but which doesn't suffer from the unphysical parameter
values of the ``Cont'' fit.

%}}} 

%{{{   rel quark model
\subsection{Relativistic Quark Model}
\label{rqm}

We now turn our attention to the derivation of $\rho(s)$ in the
relativistic quark model. In this case we can represent the number of
states as:
\begin{equation}
\mbox{Number of states} \sim \int \frac{d^3p_1}{2E_1} \int
\frac{d^3p_2}{2E_2} \delta^3(\vec{p}_1 + \vec{p}_2)
 \delta((p_1^{(0)})^2 + (p_2^{(0)})^2 - s^2),
\end{equation}
in the meson case, with an obvious generalisation for baryons. The main
difference between this case and the non-relativistic case is the energy
factors in the denominators of the normalisation. Thus we obtain two
powers of $s$ less in the meson case, and 3 powers of $s$ less in the
baryonic case. i.e.
\begin{eqnarray}
\mbox{For mesons:} & \rho_{rqm}(s) \sim s^0 \\
\mbox{For baryons:}& \rho_{rqm}(s) \sim s^2
\end{eqnarray}
Following the above analysis, the 2-point hadronic correlators can
easily be derived;
\vskip 1mm
for mesons:
\begin{equation}
G_2(t) = \frac{Z}{2M} e^{-Mt} + K \left(\frac{1}{t}\right) e^{-s_0 t}
\label{eq:rqm_mes}
\end{equation}
\vskip 1mm
and for baryons:
\begin{equation}
G_2(t) = \frac{Z}{2M} e^{-Mt} + K' \left(
\frac{1}{t^{3}} + \frac{s_0}{t^2} + \frac{s_0^2}{2t}
\right) e^{-s_0 t}.
\label{eq:rqm_bar}
\end{equation}
Note that the relativistic quark model prediction for $G_2(t)$ for
baryons is identical to the non-relativistic mesonic $G_2(t)$.
%}}} 

%{{{   Quark model fits

\subsection{Quark Model Fits}

\begin{figure}
\begin{center}
\mbox{\epsfig{file=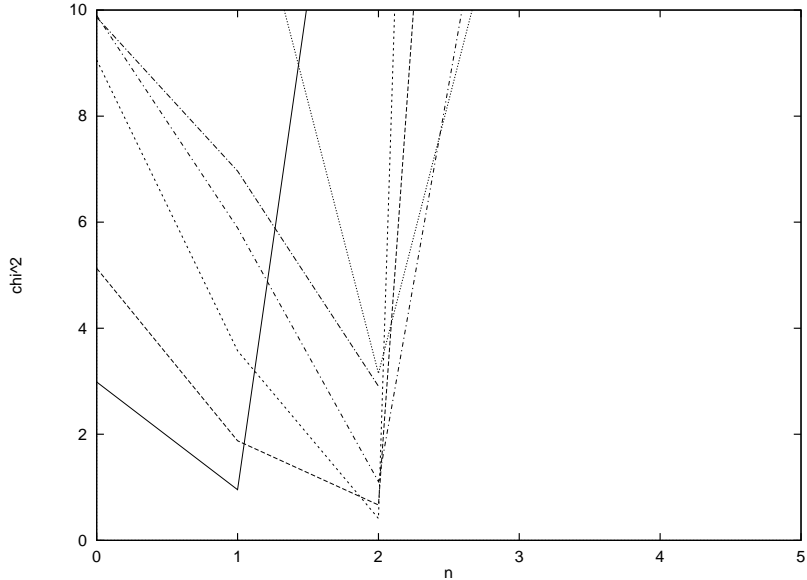, angle=270, width=11cm}}
\vskip 10mm
\caption{
\it{The $\chi^2$ value plotted against $n$ for the pseudoscalar
meson case for the fit corresponding to eq.(\protect\ref{eq:guess}).
The six lines are drawn as a guide for the eye only. They join points
from the same simulations (see Table \protect\ref{tab_latt_params}).}}
\protect\label{fig_chi2_v_n_p5}
\end{center}
\end{figure}

\begin{figure}
\begin{center}
\mbox{\epsfig{file=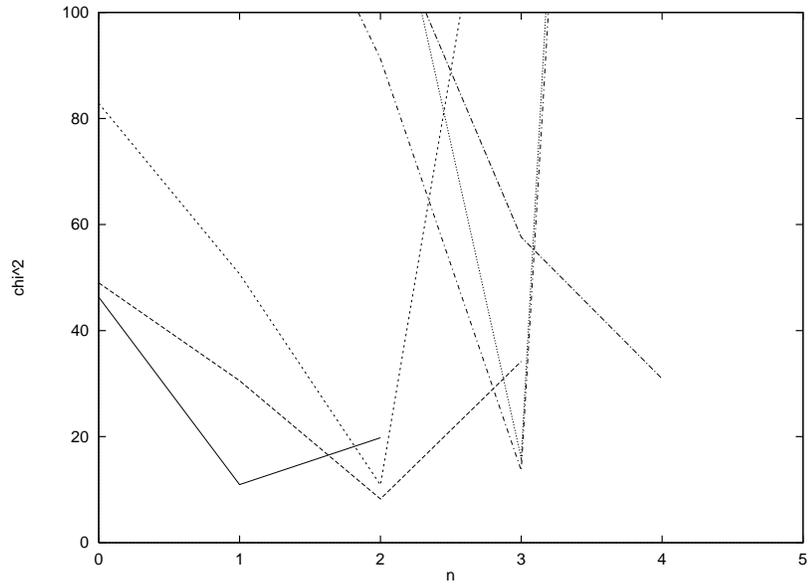, angle=270, width=11cm}}
\vskip 10mm
\caption{
\it{As in Fig. \protect\ref{fig_chi2_v_n_p5} but for the vector meson case.
No points are plotted in cases where the fits did not converge.}}
\protect\label{fig_chi2_v_n_vi}
\end{center}
\end{figure}

\begin{figure}
\begin{center}
\mbox{\epsfig{file=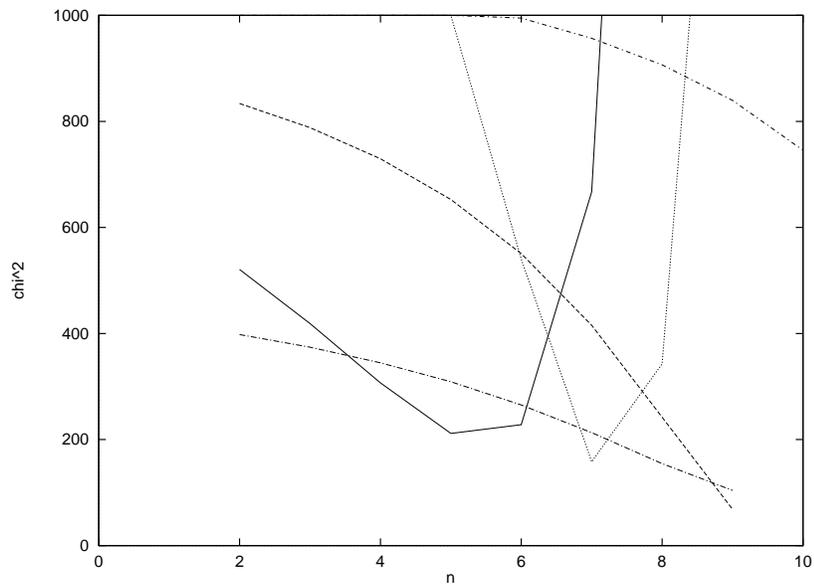, angle=270, width=11cm}}
\vskip 10mm
\caption{
\it{As in Fig. \protect\ref{fig_chi2_v_n_p5} but for the nucleon case.
No points are plotted in cases where the fits did not converge.}}
\protect\label{fig_chi2_v_n_proton}
\end{center}
\end{figure}

It is natural to wonder if there is any way of using the lattice
correlation function data to pin down the best form of the density of
states $\rho(s)$. With this in mind we assume that the density of states
has the following form:
\begin{equation}
\rho(s) = \frac{Z}{2M} \delta(s-M) + \theta(s-s_0) \; K \; s^n
\label{eq:guess}
\end{equation}
where $n$ is to be determined from the fit. The two-point correlation function
for this Ansatz is defined, as usual, from eq.(\ref{eq:spectr}).
Eq. (\ref{eq:guess}) has both the non-relativistic and relativistic forms as
special cases.
Fig. \ref{fig_chi2_v_n_p5} shows the $\chi^2$ value plotted against $n$ for
fits to the 6 sets of pseudoscalar data in table \ref{tab_latt_params}
using eq.(\ref{eq:guess}).
As can be seen from the plot, there is a distinct minimum at $n=2$ for five of
the six data sets. This confirms the non-relativistic quark model Ansatz (i.e.
$n=2$) as the preferred choice for this channel.
The vector meson case is plotted in Fig. \ref{fig_chi2_v_n_vi}. Here, as
expected, the signal is worse,\footnote
{In general, the pseudoscalar channel has the best signal to noise
ratio.}
but the minimum is certainly not far off $n=2$.
When these fits are applied to the nucleon, the data is more poorly
behaved again (see Fig. \ref{fig_chi2_v_n_proton}). 
However, it is clear that the minimum has shifted significantly to
larger $n$ values compared with the mesonic cases - 
it is now roughly between $n=5$ and $n=10$.

\begin{figure}
\begin{center}
\mbox{\epsfig{file=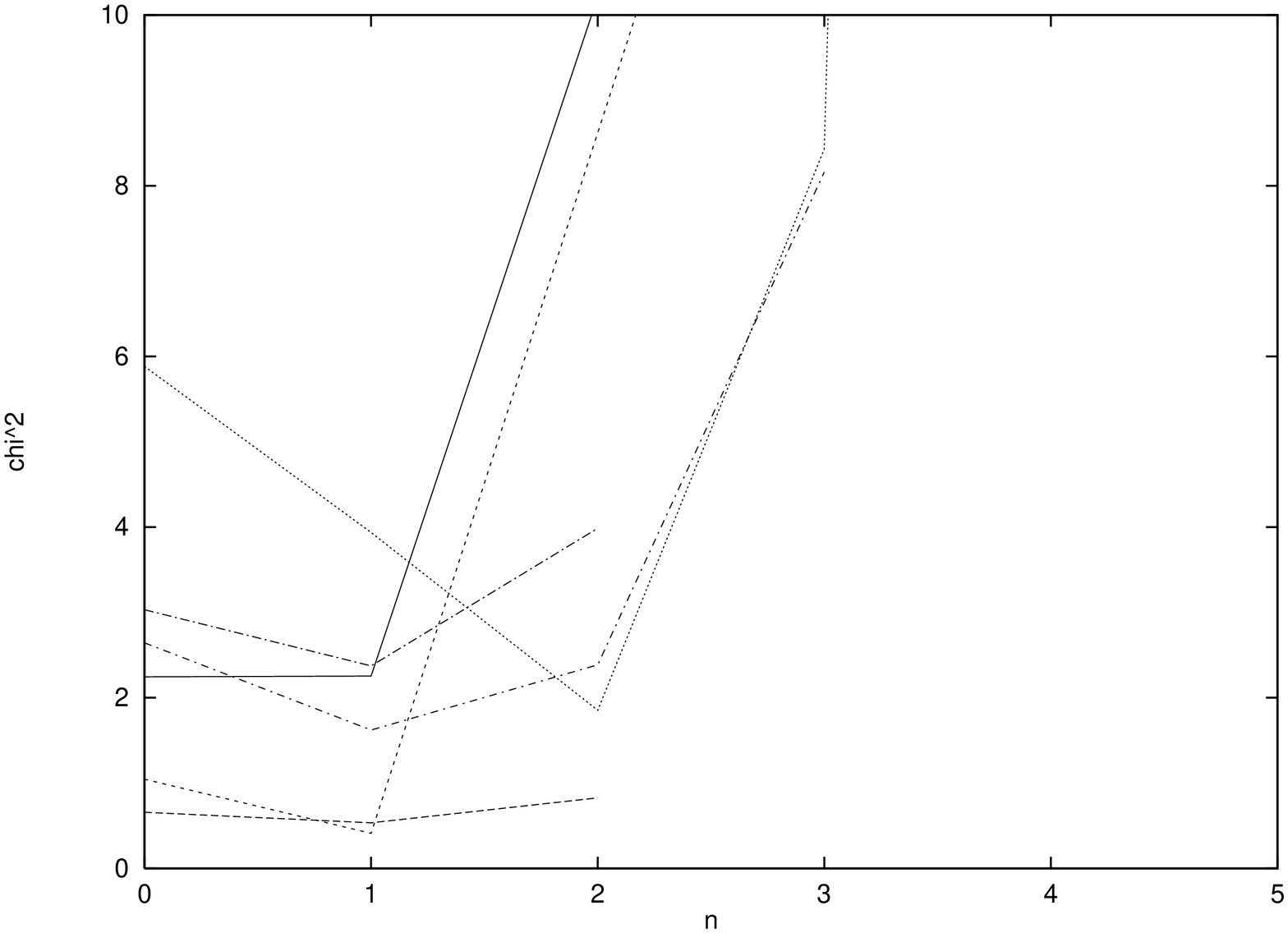, angle=270, width=11cm}}
\vskip 10mm
\caption{
\it{As in Fig. \protect\ref{fig_chi2_v_n_p5} but for lighter quark masses
corresponding to $M_V \approx 900 MeV$.
No points are plotted in cases where the fits did not converge.}}
\protect\label{fig_chi2_v_n_p5_mv900MeV}
\end{center}
\end{figure}

We now study the behaviour of the Ansatz eq.(\ref{eq:guess}) as a function
of quark mass. In Fig. \ref{fig_chi2_v_n_p5_mv900MeV} the
$\chi^2$ values for the pseudoscalar fits as a function of $n$ are
plotted for lighter quarks, corresponding to $M_V \approx 900 MeV$.
(Recall that all other data in this work is for quark masses
corresponding to $M_V \approx 1.1 GeV$ - see Table
\ref{tab_latt_params}.)
Comparing Fig. \ref{fig_chi2_v_n_p5_mv900MeV} with Fig.
\ref{fig_chi2_v_n_p5} one can see a tendency for the minimum $\chi^2$ to
decrease from $n=2$ as the quark mass decreases. This fits nicely with our
intuition - lighter quarks should eventually become relativistic which
corresponds to $n=0$ (see Sec. \ref{rqm}).

%}}} 

%{{{   fig_chi2_v_n_p5
%}}} 
%{{{   fig_chi2_v_n_vi
%}}} 
%{{{   fig_chi2_v_n_proton
%}}} 

%{{{   fig_chi2_v_n_p5_mv900MeV
%}}} 

%}}} 

%{{{   discussion & conclusion
\section{Discussion \& Conclusion}

We have begun this work with a comprehensive study of the QCD Sum Rule
Continuum Model as applied to lattice two-point correlation functions
using the method first introduced by Leinweber \cite{lw1,lw2}. We have 
extended his work by including mesonic states, by fitting lattice data
at several lattice spacings and by using two formulations of the
lattice action. We have found that the QCD Sum Rule Continuum Model
successfully fits the lattice data, and the quality of the fits is a
significant improvement over the conventional ``2-exponential'' fits.
However, while the results of the fits are successful, the values of
the parameters in the fits are unphysical. This leads us to the conclusion
(somewhat contrary to that of \cite{lw1}) that this model cannot
self-consistently fit lattice data.

We have then searched for a model that reproduces a similar functional
form as the QCD Continuum Model, but which does not have the same
restrictions on its parameters. Both the non-relativistic and
relativistic quark models were studied. By using a fitting function
which interpolated between the two, it was found that the
non-relativistic quark model was qualitatively better that its
relativistic counterpart for the quark mass studied. In the case of the
pseudoscalar meson, these remarks are more quantitative. In addition,
by studying lighter quark masses, the best fitting function was found
to tend towards the relativistic form, as expected.

Thus we have found a model which appears to correctly parametrise
the r\^ole of the excited states.
The functions that we have found can be widely used for
Lattice Gauge Theories fits when the time separation in the
correlation functions is forced to be small.
In particular, it would be interesting to apply these functions to
the case of the static case of the Heavy Quark Effective Theory on the
lattice where ground state properties are poorly isolated
(see, for example, \cite{eichten}) and in glueball studies.
%}}} 

%{{{   acknowledgements
\begin{ack}

We wish to thank the APE collaboration for allowing us to use the
lattice correlation function data presented here. We also thank Ian
Drummond, Simon Hands, Derek Leinweber, Vittorio Lubicz, 
Martin L\"uscher, Graham Shore
and John Stack for useful discussions. CRA would like to acknowledge
help from John Evans.

S. C. thanks the Organising Committee for his participation at the
16th UK Institute at Swansea, where part of this work was carried out.

This work was supported by the EC Contract ``Computational
Particle Physics'' CHRX-CT92-0051, and by the EC Human Capital
and Mobility Program, contracts ERBCHBICT941462 and ERBCHBGCT940665
and by a grant from the Nuffield Foundation.

\end{ack}
%}}} 

%{{{   bibliography

%}}} 

\end{document}